\documentclass[lettersize,journal]{IEEEtran}

\usepackage{multirow}
\usepackage{xcolor}
\usepackage{colortbl}
\usepackage{float}
\usepackage{threeparttable}
\usepackage{amsmath}
\usepackage{amsthm}
\usepackage{pifont}
\usepackage{graphicx}
\usepackage{algorithm}
\usepackage{algorithmic}
\usepackage{tablefootnote}
\usepackage{adjustbox}
\usepackage{bm}
\usepackage[most]{tcolorbox}
\usepackage{makecell}
\usepackage{array}
\usepackage{url}
\usepackage{subfig}

\newcommand{\subhead}[1]{\vspace {0.04in}\noindent{\textbf{#1.}}}

\def\inconmeasure#1{\ensuremath{vector_{#1}}}
\def\metricmeaure#1{\ensuremath{individual\, metric{#1}}}
\def\cna#1{\ensuremath{\mathtt{C}_{#1}}}
\def\nvd{\ensuremath{\mathtt{NVD}}}
\def\assessment#1#2{\ensuremath{\mathbf{v}(CVE-ID_{#1},\cna{#2})}}
\definecolor{grayblue}{RGB}{200,210,230} 

\def\cve#1{\ensuremath{{e}}}
\def\cved#1{\ensuremath{{e}_d}}
\def\othercve#1{\ensuremath{{e^*}}}

\def\assessment#1#2{\ensuremath{\mathbf{v}^{#1}\left(\cve{#2}\right)}}
\def\metric#1#2#3{\ensuremath{\mathbf{v}^{#1}_{#2}\left(\cve{#3}\right)}}
\def\textual#1#2{\ensuremath{\mathbf{t}^{#1}\left(#2\right)}}

\def\partition#1{\ensuremath{\mathcal{G}^{#1}}}
\def\inconsistent#1{\ensuremath{\mathcal{D}^{#1}}}
\def\consistent#1{\ensuremath{\mathcal{C}^{#1}}}

\newtheorem{definition}{Definition} 

\begin{document}
\title{The Cathedral and the Bazaar of Software Vulnerabilities: From the NVD to the CNAs}
\author{
Siqi Zhang,
Fabio Massacci,
and Mengyuan Zhang
\thanks{
Siqi Zhang, Fabio Massacci, and Mengyuan Zhang are with
Vrije Universiteit Amsterdam,
Amsterdam, The Netherlands.
E-mail: \{s.zhang4,f.massacci,m.zhang\}@vu.nl.
}

\thanks{
Fabio Massacci is also with
University of Trento,
Trento, Italy.
E-mail: fabio.massacci@unitn.it.
}
}



\maketitle

\begin{abstract}
For decades, the National Vulnerability Database (NVD), the ``Cathedral'', has been \emph{the} reference source for vulnerability information for downstream research and industry tasks, e.g., software update prioritization. An emerging ``Bazaar'' of diverse CVE Numbering Authorities (CNAs) has created many alternative and sometimes diverging sources. We conduct a systematic analysis of divergence in Common Vulnerability Scoring System (CVSS) metrics covering the NVD and the public CNAs. We also check for self-divergence: two identical textual descriptions of CVEs with identical CWEs are rated differently by the same CNA. The odds of diverging are widespread, not uniform and sometimes unexpected. The assessment of \textit{Attack Complexity}, \textit{User Interaction}, and \textit{Impact} are the major metrics where divergence happens. To understand the root causes, we perform a qualitative study by reaching out to the NVD and other CNAs (both open sources and proprietary products). We also discussed the findings at the CVSS Special Interest Group of FIRST, the community responsible for maintaining and evolving the CVSS standard. The key insights are that while something might be due to human errors, in some cases diverging is actually the right thing to do and might require changes in the way CVEs are generated industry-wide, in other cases explaining divergence requires access to additional FAQs. The good news is that the situation is improving since 2025, the bad news is that if one downloads the whole NVD (or another CNA dataset) from several years and uses it for predictions, the models trained on one source do not reliably generalize to a different source (accuracy can drop by 40\%). We discuss the implications for practice and research.
\end{abstract}


\section{Introduction} \label{sec:intro}

Eric Raymond \cite{DBLP:journals/firstmonday/Raymond98} introduced the word \emph{bazaar} in 1998 to positively describe distributed and collaborative software development for Linux. The code is developed by the public as opposed to the \emph{cathedral}, a single authoritative source. 


For decades, the National Vulnerability Database (NVD) has served as \emph{the} software vulnerability cathedral, the source for severity assessment. Many governmental mandates (e.g., US Executive Orders), industry regulations (e.g., Credit Cards' PCI Security), and many research papers, such as vulnerability prioritization systems~\cite{DBLP:journals/tissec/FarrisSCGJ18}, security posture monitoring tools~\cite{DBLP:conf/eurosp/JacobsRSES23}, and security metric frameworks~\cite{mell2022cvss}, rely on NVD data as a foundational input. Correctness and consistency of these data directly affect the reliability of downstream services (See detailed list in Table~\ref{tab:impacted-works-categorized}).

\begin{figure}[t]
\centering
\subfloat[\# of CNAs release CVE per year]{
  \includegraphics[width=0.24\textwidth]{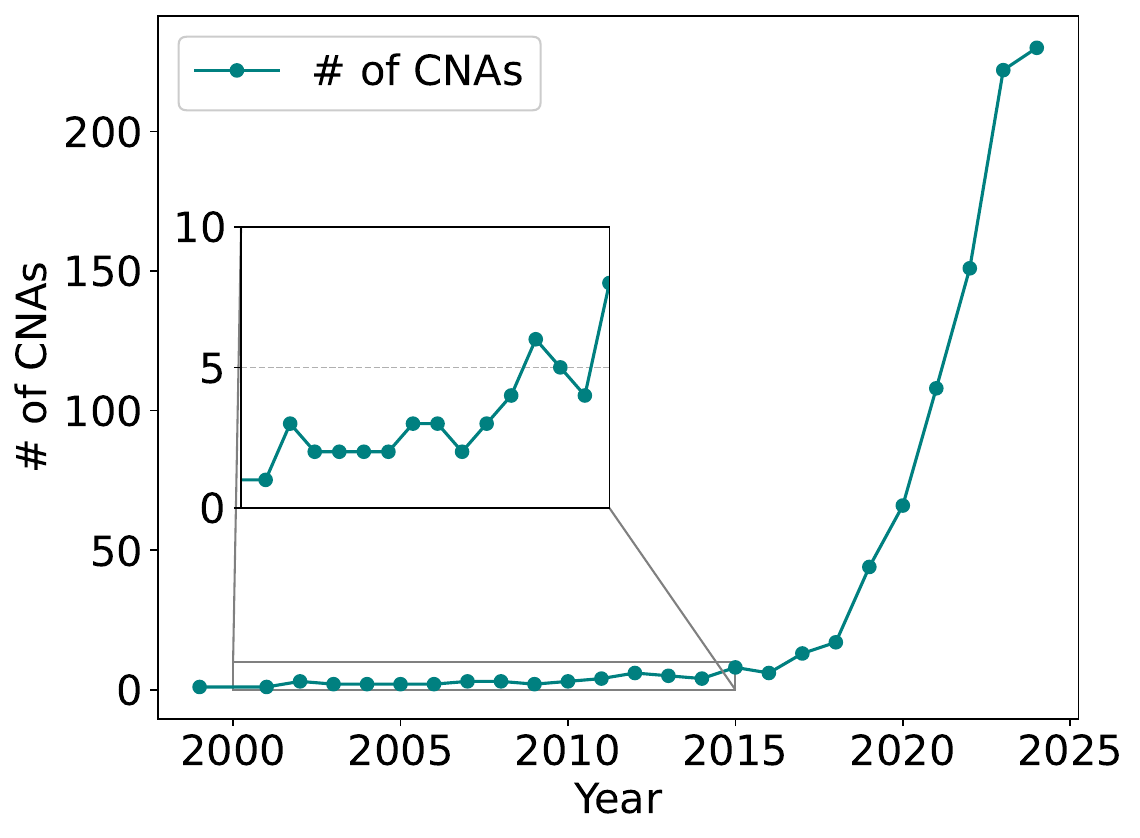}
  \label{fig:CNAyear}
}
\subfloat[Severity Levels]{
  \includegraphics[width=0.24\textwidth]{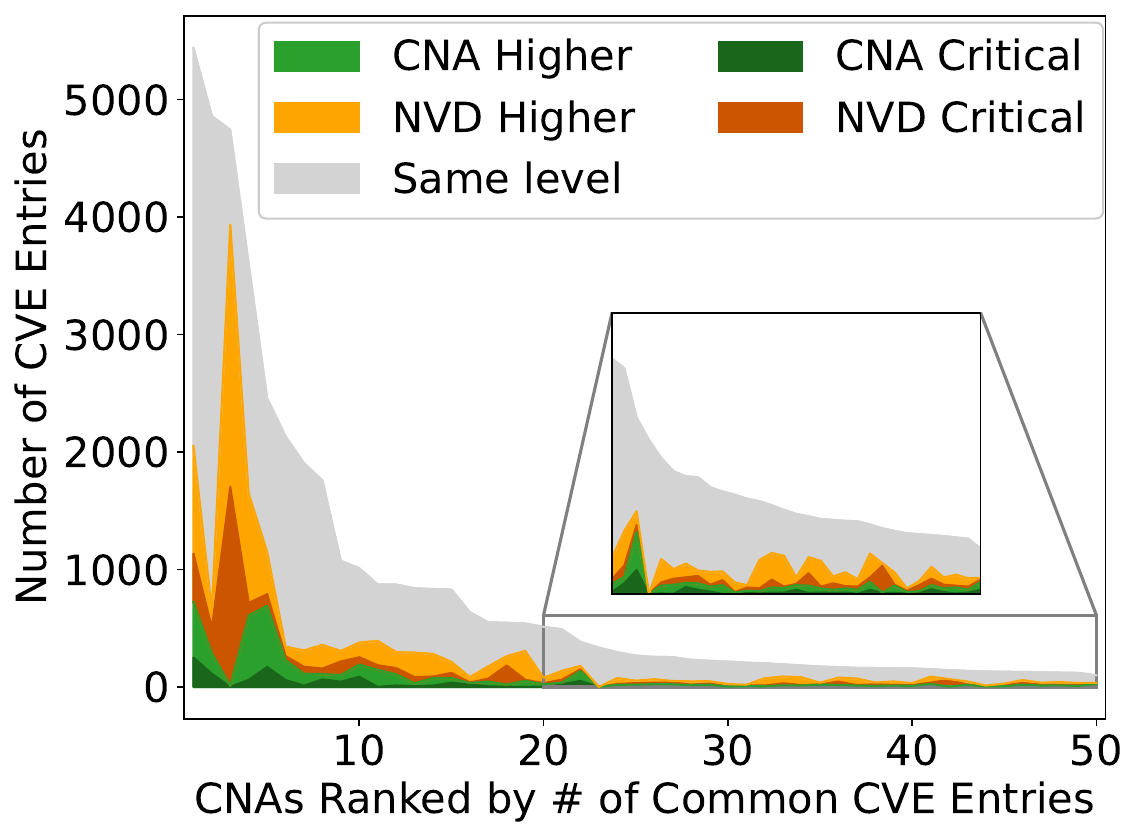}
  \label{fig:CNAcritical}
}
\caption{The growth of CNAs and severity disagreements}
\label{fig:overallCNA}
\end{figure}

Vulnerabilities boomed and, 
to scale,  the CVE program introduced the \emph{CVE Numbering Authority (CNA)} framework, authorizing registered CNAs (e.g., \textit{Microsoft}) to independently create CVE entries and assign severity assessments. Fig.~\ref{fig:CNAyear} shows the growth in the number of CNAs following the CVE program’s decision in 2016~\cite{CVE-history}. As of 2024, CNAs have issued a substantial portion of the total CVEs.
The bazaar of software vulnerability assessment was born and is now thriving.
Fig.~\ref{fig:CNAcritical} illustrates the differences in severity scored by the NVD and the severity of the \emph{same} vulnerability scored by a CNA. To decide vulnerability prioritization, researchers, companies and governments have now multiple sources. Unfortunately, they are not always consistent. 


Divergence issues within the NVD have been studied from various perspectives, including discrepancies between severity scores and vulnerability types~\cite{DBLP:journals/tdsc/AnwarACLM22}, mismatches in software names and versions~\cite{DBLP:conf/uss/DongGCXZ019}, and differences between organizational assessments~\cite{DBLP:journals/array/JiangA22}, or even within CNAs themselves in terms of CVE entries with identical~\cite{DBLP:conf/cloudcom/ZhangCZZC23}. Divergence has been also discussed in blogs~\cite{darkreading}~\cite{OpenSourceSecurity}. Coutinho et al.~\cite{DBLP:conf/ladc/CoutinhoM0LKRKL24} reveal discrepancies between aggregated CNA-assigned entries and NVD-maintained entries; their analysis was limited in scope, covering fewer than five CNAs.

A systematic and formal study comparing \emph{all} CNA-assigned entries (and not just the NVD) at a fine-grained level remains missing.  In this paper, we take the first step toward formally modeling and quantifying divergence between these sources, for the purpose of evaluating the reliability of vulnerability severity data. To understand the root causes, we perform a qualitative study by
reaching out to the NVD and other CNAs (both open sources and
proprietary products). We also discussed the findings at the CVSS
Special Interest Group (SIG) of FIRST. Our main contributions are as follows:

\begin{itemize}

\item[(i)] We introduce a unified framework to (a) quantify \emph{external} divergence: how CNA-assigned CVSS vectors differ from NVD at the vector and per-metric levels (73\% of public
CNAs have at least one diverging assessment from the NVD); and (b) assess \emph{self-divergent} by grouping CVEs with identical descriptions (description-based analysis).

\item[(ii)] We constructed an up-to-date dataset for this analysis, covering all CNAs and the NVD from 1999--2025, and augmented it with additional features (e.g., CNA type, release/update history, Common Weakness Enumeration (CWE), and Common Platform
Enumeration (CPE)). 


\item[(iii)] We uncover what \emph{type} of CNA  (e.g., a \textit{Vendor} vs. a \textit{Bug Bounty Provider}), the \textit{release history}, makes a significant difference in both the chance and the direction of divergence, and on which base metrics (Impact, Attack Complexity, and User Interaction) such divergence is most pronounced.
\item[(iv)] We provide the first empirically grounded root-cause analysis of CVSS divergence by engaging with CNAs, NVD, and the CVSS Special Interest Group (SIG). Through presentations, feedback exchanges, and interview with NVD, we identify eight root causes of divergence. The SIG is currently in the process of extending its FAQs to explicitly discuss scoring differences.
\item[(v)] We found improvements in both CVSS divergence and description quality after 2025, corresponding to industry changes in reporting. We complement these results with case studies based on implemented attacks and vulnerability discussions.
\item[(vi)] We further discuss how the impact of our findings on downstream use of the data. For example, automated, machine learning approaches that just pick up a dataset might not be learning severity but just the risk appetite of the particular CNAs or the NVD.
 \end{itemize}



\section{Terminology and Motivation} \label{sec:Preliminaries}

\subhead{Common Vulnerabilities and Exposures (CVE)} CVE is a public system for identifying and cataloging cybersecurity vulnerabilities. Each CVE entry includes a unique identifier (CVE ID), a brief description, and references to related reports and advisories. CVE records are contributed by various sources~\cite{nvd-cna-counting}.

\subhead{CVE Program} As of mid-2025, the CVE Program includes 458 CNAs across 40 countries~\cite{CVE-CNA}. The program categorizes CNAs into seven organizational types: \textit{Vendor}, manages vulnerabilities in its own products or services; \textit{Researcher}, conducts security research to identify and disclose vulnerabilities suitable for CVE tracking; \textit{Open Source}, maintains software whose source code is publicly accessible and modifiable; \textit{CERT}, Computer Emergency Response Team; \textit{Hosted Service}, refers to cloud-based services such as PaaS, IaaS, and SaaS platforms; \textit{Bug Bounty Providers}, intermediary platforms that connect vendors with researchers, often offering incentives for valid vulnerability reports; \textit{Consortium}, a group of entities joined together to work on a particular project.

\subhead{CVE Numbering Authority (CNA)} A CNA is responsible for assigning CVE IDs, publishing initial vulnerability records that describe the vulnerabilities, and may optionally include additional information such as severity scores (e.g., CVSS), affected products, and references. A special CNA, the Cybersecurity and Infrastructure Security Agency (CISA) serves as both a CNA and the only designated \emph{Authorized Data Publisher (ADP)}~\cite{CISA-ADP}.


\subhead{Common Vulnerability Scoring System(CVSS)}
The CVSS~\cite{first-cvss} is an open framework and widely recognized standard for assessing the characteristics and severity of vulnerabilities. The base metrics are typically used to calculate the severity score and determine the severity level of a given vulnerability. CVSS v3.1 includes eight base metrics: Attack Vector (\texttt{AV}), Attack Complexity (\texttt{AC}), Privileges Required (\texttt{PR}), User Interaction (\texttt{UI}), Scope (S), Confidentiality (\texttt{C}), Integrity (\texttt{I}), and Availability (\texttt{A}). Once all of them are evaluated, they are combined into a CVSS Vector using a standardized syntax. The CVSS framework computes a severity score from 0.0 to 10.0 based on base metric values, which is mapped to a qualitative level.





\begin{figure}[h!]
\centering
  \subfloat[Cross-source disagreement on the same CVE]{\includegraphics[width=.95\linewidth]{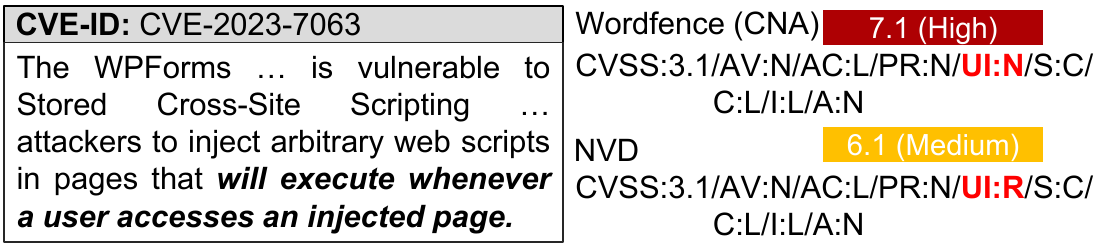} \label{fig:moti1}}
    \hspace{0.5cm}
  \subfloat[Self-divergent scores, same CNA, same description]{\includegraphics[width=.95\linewidth]{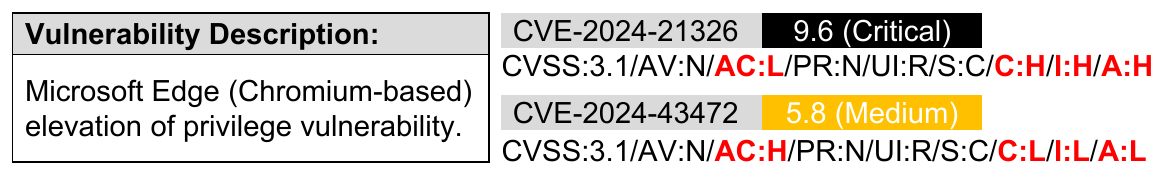}\label{fig:moti2}}
   \caption{Motivating examples of CVSS divergent scores}\label{fig:moti}
\end{figure}
\subhead{Motivating Example} 
Fig.~\ref{fig:moti} highlights two key problems: (1) divergence between CNAs and the NVD, and (2) self-divergence within identical contexts even within a single source. In Fig.~\ref{fig:moti1}, for CVE-2023-7063, \textit{Wordfence} and the NVD diverge on the \texttt{UI} base metric, resulting in different severity levels. Fig.~\ref{fig:moti2} shows that divergence arises even within individual CNA (e.g., Microsoft) itself. Two CVEs with the identical description contain four different vectors. Consequently, divergent CVSS assignments, ``Critical'' and ``Medium'', may result in different patching prioritization, thereby extending the attack window and undermining overall security. Despite the potential impact, there is currently no guidance on how to systematically quantify, interpret, or reconcile these differences.


\section{Methodology} \label{sec:methodology}




In this section, we formally model discrepancies between CNAs and the \nvd\ at two levels; \textit{Vector-level} divergency (\S\ref{sub:entry-level}) and \textit{Description-level} divergency (\S\ref{sub:group-level}). 

 
We first formally define vulnerability assessment and descriptions. We denote a \emph{CVE identifier} as \cve{i}, possibly with subscript $i,j$. We denote the corresponding \emph{vulnerability assessment by a CNA} \cna{} (e.g., CVSS base metrics values, CVSS scores, severity levels) as the function: 
\begin{eqnarray}
    \assessment{\cna{}}{i} & = & ( \metric {\cna{}} 1 i, \ldots,  \metric {\cna{}} m i)  \label{eq:vuln:assessment}
\end{eqnarray}
where \( m \) is the number of \emph{assessment dimension} of CVSS base metrics (e.g., \( m = 8 \) for CVSS v3.1) and  \metric {\cna{}} k i\ is the value of the \emph{individual base metric} of  the vulnerability assessment (e.g., for CVSS v3.1, the base metric for $k=1$ is the attack vector \texttt{AV} and the value can be \texttt{N} for network). Each CVE \cve{i} is associated by a CNA \cna\ with a \emph{textual description} \textual{\cna{}}{e}.

\subsection{Vector-level Divergency} \label{sub:entry-level}

An \cve{i}\ is considered as a divergent entry if its vulnerability assessment differs between two data sources.

\begin{definition}[Divergent Entry] \label{def:entry}
Let \cna{} be a CNA, let $\text{CVE}^{\cna{}}$ be the set of CVE entries published by \cna{}, and
let $\text{CVE}^{\nvd}$ be the set of CVE entries reported by the \nvd. 
A CVE entry $\cve{i} \in \text{CVE}^{\cna{}}$ is a \emph{divergent entry} if there exists an entry $\cve{i} \in \text{CVE}^{\nvd}$ such that the vulnerability assessment differs \( \assessment{\cna{}}{i} \neq \assessment{\nvd}{i} \).
\end{definition}


We propose two complementary metrics to quantify divergent degrees: the \textit{\inconmeasure{}-Divergency}, and the \textit{\metricmeaure{}-Divergency}, which further breaks down Divergency across individual base metrics. The first \inconmeasure{}-Divergency metric calculates the distance between two CVSS vectors by adopting the concept of \textit{Hamming distance}, originally introduced to measure the number of bit errors in message transmission over a noisy channel. In our context, each base metric (e.g., Attack Vector) contributes 1 to the total distance if the values differ, otherwise 0.

\begin{definition}[\inconmeasure{}-Divergency] \label{eq:kd}
Let \cved{} be a divergent entry between the CNA \cna{} and the \nvd\ then, the \inconmeasure{} \emph{divergent metric} is defined as the Hamming distance between the two assessments with $m$ individual metrics:
\begin{eqnarray}\label{hamming}
\inconmeasure{\cved{}} & = & \sum_{k=1}^m \mathbf{1}_{\metric{\cna{}}{k}{i} \neq \metric{\nvd}{k}{i}}
\end{eqnarray}
\end{definition}


Since Definition~\ref{eq:kd} does not reveal which specific base metrics are more prone to diverge. We therefore propose the \metricmeaure{}-Divergency to capture this aspect.

\begin{definition}[\metricmeaure{}-Divergency]\label{def:permetric}
Let \cna\ be a CNA, let $\text{CVE}^{\cna{}}$ be the set of CVE entries published by \cna{}, and $\text{CVE}^{\nvd}$ be the set of CVE entries reported by the \nvd, and let $\mathcal{D} =  \text{CVE}^{\cna{}} \cap \text{CVE}^{\nvd}$, the \metricmeaure{}-Divergency between the CNA \cna\ and the \nvd\ on the assessment metric \( k \in \{1, \ldots, m\} \) is defined as:

\begin{eqnarray}\label{eq:permetric}
\metricmeaure{}_k(\cna{}) & = &  \sum_{\cve{i} \in \mathcal{D}} \mathbf{1}_{\metric{\cna{}}{k}{i} \neq \metric{\nvd}{k}{i}}
\end{eqnarray}

\end{definition}
The \metricmeaure{}-Divergency represents absolute difference of overlapping CVE entries for which CNA \cna{} and the \nvd\ disagree on the \( k \)-th CVSS metric, and $\frac{1}{\mid\mathcal{D}\mid}*\metricmeaure{}_k(\cna{})$ presents the ratio.



\subsection{Description-level Divergency} \label{sub:group-level}

After studying the divergences for individual CVE entries rated by different sources, we address the second issue mentioned in the motivating example, i.e., how identical descriptions are rated in both CNAs and the \nvd. The definitions are based on CNAs, however, the \nvd\ shares the same concept except changing the notations to the $*^{\nvd}$.

\begin{definition}[Textual Equivalence]
Let \cna\ be a CNA and $\text{CVE}^{\cna{}}$ be the set of CVE entries published by \cna{}, an equivalence relation \( \sim_{\cna{}} \) over $\text{CVE}^{\cna{}}$ is induced by the textual description of \cna{} as follows
\begin{eqnarray}
    \cve{i} \sim_{\cna{}} \othercve{j} &\text{if and only if}& \textual{\cna{}}{\cve{i}} = \textual{\cna{}}{\othercve{j}}
\end{eqnarray}
\end{definition}
This relation partitions $\text{CVE}^{\cna{}}$ into equivalence classes \( \partition{\cna{}} = \{G_1, G_2, \ldots, G_k\} \), where each group \( G \in \partition{\cna{}} \) contains entries that share an identical textual description.
A group can however be inconsistent due to internal disagreement in the assessment of identically described entries.

\begin{definition}[Self-Divergent Group]
\label{def: inconsistent}
Let \cna\ be a CNA,  
let $\text{CVE}^{\cna{}}$ be the set of CVE entries by \cna{}, and 
let $\mathcal{G}^{\cna{}}$be the partition $\sim_{\cna{}}$ induced by the identical textual description from the CNA \cna{}, 
a group \( G \in \mathcal{G}^{\cna{}} \) is \emph{self-divergent} if there exist two entries $\cve{i}, \othercve{j}\in G$ such that their assigned metric vectors differ, $\assessment{\cna{}}{i}\neq \mathbf{v}^{\cna{}}(e^*)$.
\end{definition}

We denote the set of all divergent description groups for a CNA \cna{} as \inconsistent{\cna{}}, we also define the consistent group as the set 
of CVEs which has at least two CVEs but where all descriptions have the same text and the same CVSS metric vector \consistent{\cna{}}.

\setlength{\arraycolsep}{2pt}
{\small
\begin{eqnarray}
    \inconsistent{\cna{}} & = & 
    \left\{G \in\partition{\cna{}} \mid 
    \exists \cve{i}, \othercve{j} \in G: 
    \cve{i} \neq \othercve{j} \wedge  
    \assessment{\cna{}}{i} \neq  \mathbf{v}^{\cna{}}(e^*) 
    \right\} \\
    \consistent{\cna{}} & = & 
    \left\{G \in\partition{\cna{}} \mid 
    \exists \cve{i}, \othercve{j} \in G: 
    \cve{i} \neq \othercve{j} \wedge 
    G \not\in \inconsistent{\cna{}} 
    \right\}
\end{eqnarray}}

\begin{definition}[Odds Ratio of the Group]
\label{def:odds-ratio} Let $\sim_{\cna{}}$ and $\sim_{\nvd}$ be the equivalence relations induced by the textual descriptions provided by the CNA \cna{} and the \nvd, \emph{divergency odds ratio} for a CNA \( \cna{} \) relative to the \nvd\ is then defined as:
\begin{eqnarray}
\label{eq: odds-ratio}
\text{OR}^{\cna{}/\nvd} = 
\frac{
\left| \inconsistent{\cna{}} \right| / \left| \consistent{\cna{}} \right|
}{
\left|\inconsistent{\nvd} \right| / \left| \consistent{\nvd} \right|
}
\end{eqnarray}

\end{definition}

An OR greater than 1 indicates that the CNA exhibits a higher rate of divergency relative to the \nvd, while an OR less than 1 suggests otherwise.

\subsection{Root Cause Qualitative Analysis}
\label{subsec:rootCause}

To understand the root causes of divergence, we also perform a qualitative analysis. Since the CVSS Special Interest Group (SIG) has recently discussed CVSS inconsistent scoring practices, we contacted the SIG chair, several CNA representatives, and the NVD with divergent cases from our analysis. To preserve confidentiality, we anonymize the identities and organizational affiliations of the contacted CNA representatives throughout the paper.

We contacted the NVD and six CNA representatives who had either participated in the CVSS SIG discussions or whose organizations exhibited significant divergence in our analysis. We informed them that we were investigating the issues discussed during the meeting, shared representative observations from our preliminary analysis, and asked whether they would be interested in reviewing our findings. Before presenting our work at the subsequent SIG meeting, we offered to share both the presentation slides and the CNA-specific analysis so that their comments could be incorporated into the discussion. The invitation email is provided in Appendix Fig.~\ref{fig:first_email}.

After the NVD and the CNA representatives expressed interest in our findings, we sent a follow-up email containing our initial presentation slides and CNA-specific files containing the associated CVE cases. The slides summarized our preliminary findings presented in Section~\ref{sub:entry-level-analysis}, together with illustrative examples and the corresponding analysis results. Based on these materials, the participants reviewed our findings and provided feedback on the possible reasons for the observed inconsistencies. Several explanations were independently raised by multiple organizations, which we grouped into eight recurring root causes (RCs). We then incorporated these preliminary RCs into the presentation and discussed them at the subsequent CVSS SIG meeting, where an additional seven experts provided further comments and refinements. The follow-up email is provided in Appendix Fig.~\ref{fig:second_email}.






\begin{table*}[t!]
\caption{Dataset Summary of Pairwise and Consumer-View (1999-2025.01)}
\label{tab:dataset}
\centering
\resizebox{0.8\linewidth}{!}{
\begin{tabular}{c|c|c|c|c|c|c} 
\hline
\bfseries{Analysis Setting} & \bfseries{\# CVE entries} & \bfseries{\# CNAs} & \bfseries{\# CNAs overlap \nvd} & \bfseries{Comparable CVEs} & \bfseries{20 CNAs pairs} & \bfseries{Change history} \\ 
\hline \hline
Pairwise & \multirow{2}{*}{191,009} & \multirow{2}{*}{288} & 266 & 44,180 & 33,831 & 44,123 \\ \cline{4-7}
Consumer-View &  &  & 288 & 72,122 & 58,249 & -- \\ 
\hline
\end{tabular}}
\end{table*}

\begin{table}
\caption{CNA Organization Types}
\label{tab:CNA_or}
\centering
\resizebox{\linewidth}{!}{
\begin{tabular}{c|c|c|c|c|c|c} 
\hline
\bfseries{Organization Types}  & \bfseries{V}  & \bfseries{R} & \bfseries{OS}  & \bfseries{CERT}  & \bfseries{HS} &\bfseries{BBP}   \\ \hline
\bfseries{\#CNA} & 214 & 61 & 68 & 14 & 13 & 4 \\
\hline
\bfseries{\#CVE} & 29,821 & 20,896 & 15,961 & 2,519 & 6,060 & 3,692 \\
\hline
\end{tabular}}
\footnotesize{
\raggedright 
\scriptsize{\textbf{Note:} V = Vendor; R = Researcher;  OS = Open Source; CERT = Computer Emergency Response Team; HS = Hosted Service; BBP = Bug Bounty Provider
}}
\end{table}

\section{Dataset} \label{subsec:data}

We compile the most complete CNA-NVD dataset (available in the Artifact) by combining all available CVEs from VulnCheck~\cite{vulncheck} (covering NVD~\cite{NVD} and MITRE CVE List~\cite{CVE-Project}). It enables large-scale, reproducible analysis of cross-source CVSS divergencies.

\subhead{Data Pre-processing} We excluded entries whose descriptions begin with ``UNSUPPORTED WHEN ASSIGNED'', as this tag indicates that all affected products or versions were already end-of-life (EOL) or no longer supported by the vendor when the CVE was assigned, and therefore fall outside the scope of our study. The resulting dataset contains 191,009 CVEs with CVSS v3.1 vectors as of January 2025, including 72,122 CNA entries and 118,887 NVD entries. 

To ensure consistent organization-level analysis, we further normalize CNA identities.
When an organization is registered as a \emph{single} CNA, we consolidate all entries associated with that organization (e.g., \texttt{ykramarz@cisco.com} and \texttt{psirt@cisco.com} are mapped to \textit{Cisco}). When an organization operates \emph{multiple} CNAs for different product lines or programs, we keep them separate, e.g., \textit{Samsung Mobile} and \textit{Samsung TV} are treated as distinct CNAs, and CISA is represented by three CNAs: \textit{CISA}, \textit{ICS{-}CERT}, and \textit{CISA{-}ADP}. 

The entries containing scores from both CNA and NVD are denoted as  \textit{Pairwise} setting. According to our interviews with NVD analysts, a CNA-only CVSS score may be perceived as agreed upon by NVD when no alternative NVD score is provided, and the entry has ``Analyzed'' status. Since consumers often retrieve CVSS data and use it directly, we treat all CNA-only CVSS scores as mutually agreed-upon entries, i.e., with \inconmeasure{}-Divergency = 0. We denote this assumption as the \textit{Consumer-View} setting. In total, we obtained 44,180 CVEs under the \textit{Pairwise} setting, i.e., CVEs rated by both NVD and a CNA, and 72,122 CVEs under the \textit{Consumer-View} setting, which additionally includes CNA-only cases by assuming that NVD assigns the same CVSS vector.

\subhead{Root Cause Feature Collection} To investigate potential \emph{drivers} of CNA–NVD divergent, we first annotate each CVE with the CNA’s organization type~\cite{CNAs} (\textit{Vendor}, \textit{Researcher}, \textit{Open Source}, \textit{CERT}, \textit{Hosted Service}, \textit{Bug Bounty Provider} detailed in Table~\ref{tab:CNA_or}). To analyze temporal precedence (who acts first) and follow-on behavior, we reconstruct scoring timelines from JSON metadata exposed by both the NVD and the CVE Program. Specifically, we use the NVD CVE Change History API\footnote{\url{https://services.nvd.nist.gov/rest/json/cvehistory/2.0}} and the CVE Program API\footnote{\url{https://cveawg.mitre.org/api/cve/}}. The NVD API provides event-level histories with action timestamps and sources; the CVE Program API exposes CNA-side metadata.

By aligning timestamps across the two APIs, we derive (i) the \emph{first appearance} of a CVSS v3.1 assessment per CVE and source (``who-first''), and (ii) an complete, ordered sequence of subsequent updates. We then study whether, and within what horizon, one source \emph{follows} another, conditioning on CNA type and text-quality features. In totally, we collected 44,123 CVE entries.

\section{Divergence Analysis} \label{sub:entry-level-analysis}
\label{sub:entry-level-analysis}

In this section, we evaluate \emph{vector-level} divergency between CNA and NVD assessments using both \inconmeasure{} and 
\metricmeaure{} divergency in Section~\ref{subsec:vector-level}, then we continue with description-level in Section~\ref{sub:group-level-analysis}. 
Because CNA coverage varies widely (from $\sim\!10^2$ to $\sim\!10^4$ CVEs), we report \metricmeaure{}–Divergency as a \textbf{percentage} to enable fair comparisons across CNAs; the corresponding \textbf{absolute counts} are easily recoverable (provided in the Appendix Table~\ref{tab:metric_level_disagreement}).

\subsection{Vector-level Divergency} \label{subsec:vector-level}

\begin{figure}[t]
\centering
\includegraphics[width=0.6\linewidth]{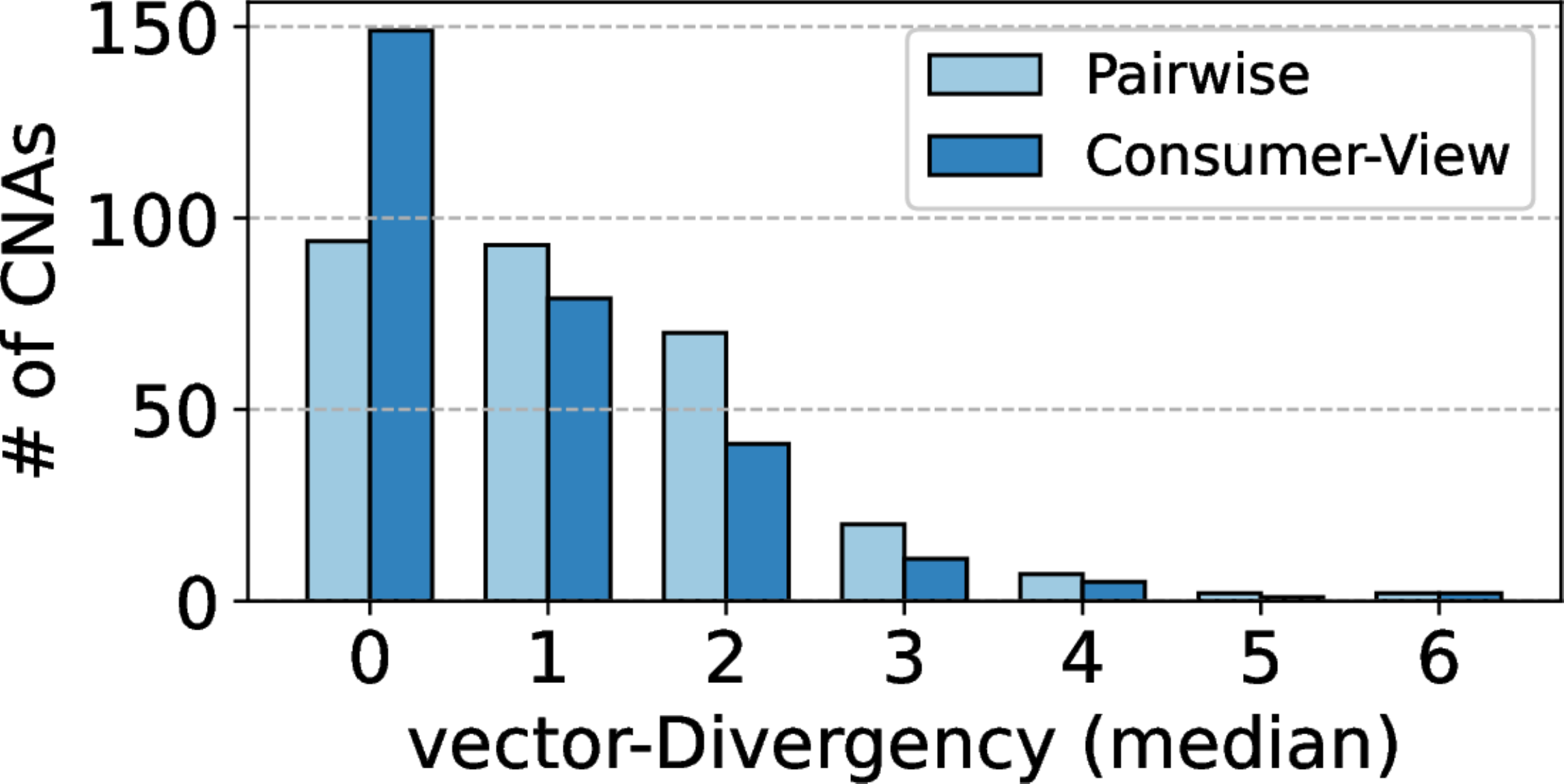}
\caption{Distribution of CNAs by \inconmeasure{}-Disagreement values under the Pairwise and Consumer-View settings}
\label{fig:dMedium_266CNA}
\end{figure}

\subhead{Overall Analysis}  Fig.~\ref{fig:dMedium_266CNA} summarizes CNA–NVD disagreement using \inconmeasure{}-Divergency. Both \textit{Pairwise} and \textit{Consumer-View} settings show \textbf{194/266 (73\%)} and \textbf{139/288 (48\%)} of CNAs have a \emph{median} of at least \(1\) \inconmeasure{}  divergent, which suggests divergent is fairly common between CNAs and NVD in both views.


\begin{figure}[h]
\centering
  \subfloat[Distribution of \inconmeasure{}-level divergent values]{\includegraphics[width=.95\linewidth]{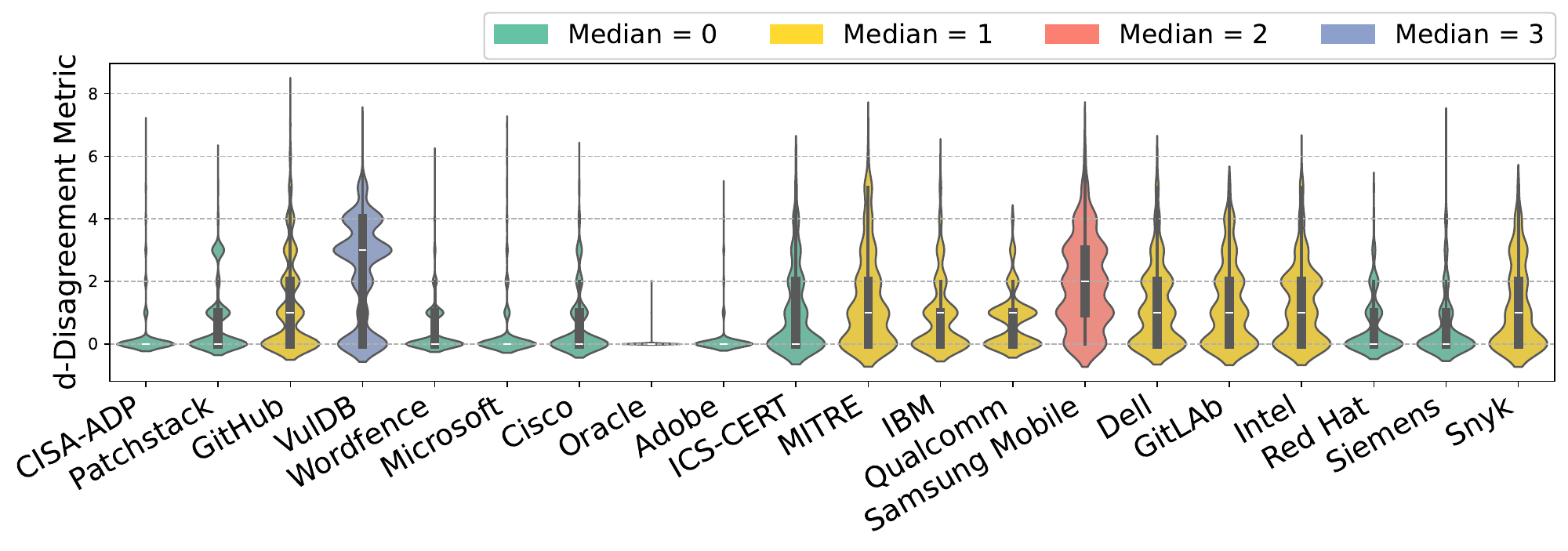} \label{fig:violin_plot}}
    \hspace{0.5cm}
  \subfloat[Proportion of \metricmeaure{}-Divergency]{\includegraphics[width=.95\linewidth]{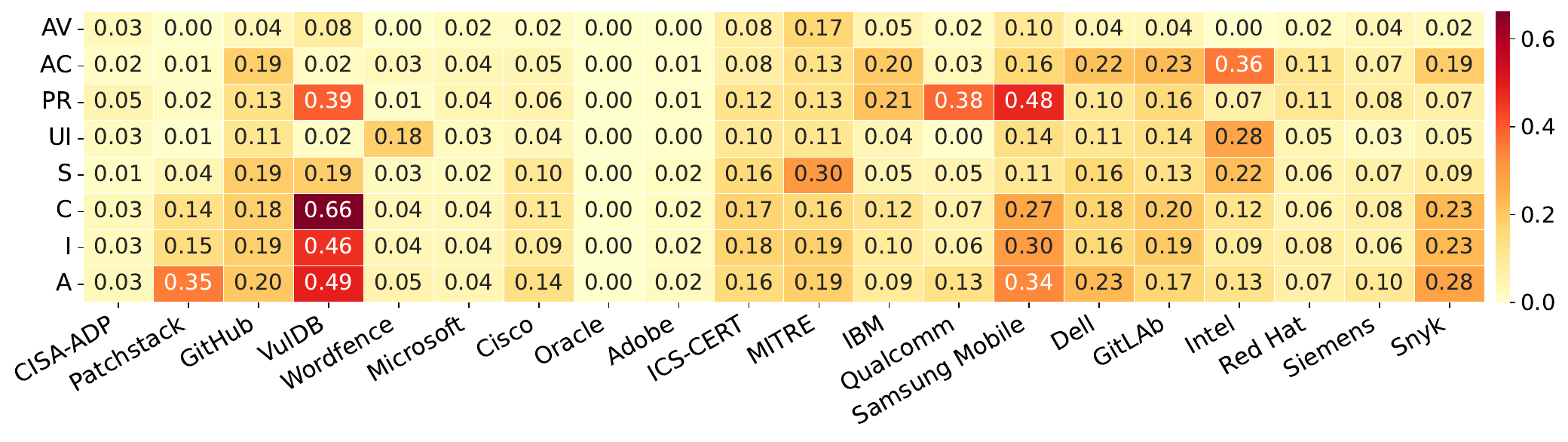}\label{fig:heatmap}}
   \caption{Top 20 CNAs compared to the NVD }\label{fig:dismetric}
\end{figure}

\subhead{\inconmeasure{}-Divergency Analysis}
\label{subsub:hamming}
We compute per CVE vector distance for every CNA–NVD pair that under \textit{Consumer-View} perspective. We then examine its distribution for the selected CNAs to characterize their divergent patterns.
Fig.~\ref{fig:violin_plot} summarizes the results: each violin shows the density of distances (wider segments indicate higher mass), with the median marked (color/shade) and interquartile range overlaid. This view highlights both \emph{where} disagreement typically concentrates and \emph{how} it varies across CNAs (spread, skew, and outliers). For readability, we only provide the Top 20 CNAs, which account for \textbf{81\%} of the data. Full results for all CNAs are included in the artifacts.



All the top 20 CNAs contain divergent entries. \textit{VulDB} shows the highest divergence, with a median value of \(3\), followed by \textit{Samsung Mobile} with a median value of \(2\). In contrast, \textit{Oracle} shows the strongest agreement with the NVD.

\begin{tcolorbox}[
    colback=gray!10,    
    colframe=black,    
    boxrule=0.4pt,     
    arc=1pt,           
    left=4pt, right=4pt, top=2pt, bottom=2pt, 
    enhanced,
    sharp corners
]
\textbf{Takeaway 1.} Even a single metric difference may change the overall
severity level and impact patch prioritization. Divergency is substantially higher under the \textit{Pairwise} setting (73\%) than under \textit{Consumer-View} (48\%), and remains common across major CNAs, with half of the top 20 exhibiting median disagreement \( \geq 1 \). 
\end{tcolorbox}

\subhead{\metricmeaure{}-Divergency Analysis}\label{subsub:heatmap} 
The \inconmeasure{}-Divergency provides a high-level overview of discrepancies between CNAs and the NVD. For more granular analysis, we use \metricmeaure{}-Divergency to identify which CVSS base metrics drive the observed divergent. 

The notable divergent \metricmeaure{}s in \textit{VulDB} are \texttt{Impact} metrics (\texttt{C/I/A}) as shown in Fig.~\ref{fig:heatmap}. \texttt{PR} metric is also the main cause for such a divergence. These per-metric patterns explain the concentration at \(\inconmeasure{}=3\) and \(\inconmeasure{}=4\) in Fig.~\ref{fig:violin_plot}: divergences in \texttt{Impact} and \texttt{PR} are major contributors to the overall divergent. \textit{Samsung Mobile} and \textit{Snyk} show similarly elevated divergency across all \texttt{Impact} metrics. \textit{Samsung Mobile} also exhibits relatively high divergency in the \texttt{PR} metric, and similar patterns are observed for CNAs such as \textit{Qualcomm}. As a conclusion, the impact-related metrics (\texttt{C/I/A}) and \texttt{PR} are common sources of disagreement across Top 20 CNAs.

\begin{tcolorbox}[
    colback=gray!10,    
    colframe=black,    
    boxrule=0.4pt,     
    arc=1pt,           
    left=4pt, right=4pt, top=2pt, bottom=2pt, 
    enhanced,
    sharp corners
]
\textbf{Takeaway 2.} Certain metrics (e.g., PR and Impact) show higher divergence rates across top 20 CNAs, suggesting inherent ambiguity in these metrics and the need for clearer internal agreement for specific CNAs. 
\end{tcolorbox}

\subsection{Description-Level Divergency}
\label{sub:group-level-analysis}


To analyze description-level divergence, we first match descriptions that appear in both a CNA and the NVD. For any description with at least two CVEs in a source, we form a \emph{group} (per source). A group is considered \emph{convergent} if all CVSS vectors are identical; otherwise, \emph{divergent}.

During the interview, CNA analysts explained that identical descriptions may result from the description templates. Further augment with Common Weakness Enumeration (CWE) and/or affected platforms, as represented by Common Platform Enumeration (CPE) identifiers, may help explain the observed divergence. Thus, we refine our grouping by additionally requiring that CVEs within a group share the same CWE and CPE attributes. Specifically, refined groups are defined by the tuple of description, CWE, CPE vendor, and CPE product. CWE values are grouped by exact match, with missing values treated as a distinct category and handled equivalently to other CWE values during grouping. Table~\ref{tab:cna_or_entropy} compares the self-divergence of CNA and NVD under both the \textit{Description-only} and refined \textit{Description+CWE+CPE} settings. \footnote{For statistical reliability, we focus on the 7 CNAs that have \(\ge\!10\) comparable groups under both \textit{Description-only} and \textit{Description+CWE+CPE} settings, ensuring sufficiently stable comparisons, as small sample sizes can lead to unreliable estimates~\cite{williams2025samplesize,schonbrodt2013sample}.}

\subhead{Odds Ratio Analysis} The odds ratio is calculated according to Equation~\ref{eq: odds-ratio}. 
Table~\ref{tab:cna_or_entropy} compares the self-divergence of CNA and NVD under both the \textit{Description-only} and the refined \textit{Description+CWE+CPE} settings. Overall, incorporating CWE and CPE significantly reduces the number of inconsistent groups for both CNA and NVD across all selected sources. For example, \textit{Microsoft} shows a substantial decrease in divergent groups from 207 to 75 (CNA) and from 160 to 62 (NVD), indicating that a large portion of the apparent inconsistency under description-only grouping can be attributed to differences in vulnerability types or affected products.

Despite this reduction, the relative divergency between CNA and NVD, as reflected by the odds ratio (OR), remains largely stable for most sources. OR values greater than 1 indicate higher divergencies in CNA compared to NVD, vice versa. For instance, \textit{Microsoft} (2.02 \(\rightarrow\) 1.40) and \textit{CISA-ADP} (6.64 \(\rightarrow\) 3.5) continue to exhibit higher divergency in CNA compared to NVD, while \textit{Cisco} (0.08 \(\rightarrow\) 0.38) remains more converged than the NVD.  \textit{Dell} shows near-perfect convergent within these refined groups. And \textit{Oracle} and \textit{Adobe} show no difference between the two settings.

\begin{tcolorbox}[
    colback=gray!10,    
    colframe=black,    
    boxrule=0.4pt,     
    arc=1pt,           
    left=4pt, right=4pt, top=2pt, bottom=2pt, 
    enhanced,
    sharp corners
]
\textbf{Takeaway 3.} Refining grouping from \textit{Description-only} to \textit{Description+CWE+CPE} substantially reduces divergent groups, while the relative consistency between CNAs and NVD remains largely stable. This highlights the need to include additional attributes to identify internal consistency; odds ratios help identify more consistent sources for patch prioritization.  
\end{tcolorbox}

\begin{table}
\caption{Description-level analysis of self-divergence and convergent groups between selected CNAs and NVD. Values are shown as Description-only (Description+CWE+CPE).}
\label{tab:cna_or_entropy}
\centering
\footnotesize
\renewcommand{\arraystretch}{1.25}
\setlength{\tabcolsep}{3pt}
\begin{tabular}{
  >{\centering\arraybackslash}p{1.7cm}|
  >{\centering\arraybackslash}p{1cm}|
  >{\centering\arraybackslash}p{1cm}|
  >{\centering\arraybackslash}p{1cm}|
  >{\centering\arraybackslash}p{1cm}|
  >{\centering\arraybackslash}p{1.5cm}
}
\hline
Source & 
\(| \inconsistent{\text{C}} |\) &
\(| \consistent{\text{C}} |\) &
\(| \inconsistent{\text{NVD}} |\) &
\(| \consistent{\text{NVD}} |\) &
$\text{OR}^{C/NVD}$ \\
\hline

& \multicolumn{5}{c}{Description-only (Description+CWE+CPE)}\\
\hline

Microsoft & 207 (75) & 84 (84) & 160 (62) & 131 (97) & 2.02 (1.40) \\
CISA-ADP & 28 (9) & 52 (36) & 6 (3) & 74 (42) & 6.64 (3.5) \\
OpenHarmony & 3 (2) & 10 (8) & 2 (1) & 11 (9) & 1.65 (2.25)  \\
Dell & 9 (3) & 19 (11) & 6 (0) & 22 (14) & 1.74 (U) \\
Adobe & 7 (1) & 73 (11) & 7 (1) & 73 (11) & 1 (1) \\
Cisco & 8 (4) & 130 (37) & 61 (9) & 77 (32) & 0.08 (0.38) \\
Oracle & 0 (0) & 27 (20) & 0 (0) & 27 (20) & U (U) \\

\hline
\end{tabular}

\raggedright
\footnotesize{
\textit{Note:} \textbf{U} denotes undefined odds ratios, which occur when the denominator in the odds ratio computation is zero.
}

\end{table}

\section{Key Factors in Divergency} \label{sec:keyfactors}

We conducted an experiment to investigate which factors influence divergent vulnerability ratings, particularly focusing on \inconmeasure{}-level and \metricmeaure{}-level divergences.


\begin{table}[t]
\caption{Logistic and Linear Regression Results}
\label{tab:RegressionNew}
\centering
\footnotesize
\begin{tabular*}{\columnwidth}{@{\extracolsep{\fill}}l|cc|cc}
\hline
\multirow{2}{*}{\textbf{Selected Factors}} 
 & \multicolumn{2}{c|}{\textbf{Log.\ $NVD \neq CNA$}} 
 & \multicolumn{2}{c}{\textbf{Lin.\ $CNA - NVD$}} \\
 & Coef. & Sig. & Coef. & Sig. \\
\hline
Constant & -1.157 & \textbf{***} & -1.477 & \textbf{***} \\ \hline

Vendor & 0.653 & \textbf{***} & 1.187 & \textbf{***} \\

Open Source & -0.265 & \textbf{***} & 0.091 &  \\

Researcher & 2.286 & \textbf{***} & -1.108 & \textbf{***} \\

Bug Bounty Provider & 0.495 &  & 0.055 & \\

Hosted Service & -1.544 & \textbf{***} & 0.445 & \textbf{**} \\

CERT & 0.577 & \textbf{***} & 1.221 & \textbf{***} \\
\hline

Who First & 2.072 & \textbf{***} & -0.010 & \\

CWE & 0.159 & \textbf{***} & -0.036 &   \\

Description Length & -0.002 & \textbf{***} & 0.002 & \textbf{***} \\


\hline

Who First $\times$ V & -1.080 & \textbf{***} & $-2.0 \times 10^{-4}$ & \\

Who First $\times$ OS & 0.211 & \textbf{***} & -0.362 & \textbf{***} \\

Who First $\times$ R & -1.867 & \textbf{***} & 0.770 & \textbf{***} \\

Who First $\times$ BBP & 1.935 & \textbf{***} & 0.191 & \\

Who First $\times$ HS & 0.383 & \textbf{**} & -0.427 & \textbf{**} \\

Who First $\times$ CERT & -1.266 & \textbf{***} & -0.081 & \\
\hline
\end{tabular*}

\footnotesize
\raggedright
\textbf{Significance levels:}
$^{*} p < 0.05$, 
$^{**} p < 0.01$, 
$^{***} p < 0.001$.
Non-significant results ($p \ge 0.05$) are left blank. \\

\textbf{Note:}
Logistic regression uses 43,997 entries (dependent variable 
$Vector_{NVD}(cve)\neq Vector_{CNA}(cve)$), pseudo-$R^2 = 0.117$.
Linear regression uses only the 26,931 conflicting entries 
(dependent variable $Score_{CNA}(cve)-Score_{NVD}(cve)$), 
$R^2 = 0.118$, $F(15, 26{,}931)=240.5$.
\end{table}

\subsection{Factor Selection and Analysis} 
\label{subsec:appearance}
We first applied a logistic regression analysis to identify the factors influencing scoring divergency, followed by a linear regression to examine the magnitude and direction of how these factors affect the CVSS base score differences under the \textit{Pairwise} setting. The dependent variable of the logistic regression indicates whether a CNA-assigned score differs from that of the NVD (\inconmeasure{}-Divergency $>0$), while the dependent variable of the linear regression represents the numerical difference in CVSS base scores between the CNA and the NVD, considering only conflicting CVE entries (Hamming distance $>0$ and CNA-assigned score $\neq$ NVD-assigned score). Both analyses share the same independent variables, including six CNA organization types~\cite{CNAs} (\textit{Vendor}, \textit{Researcher}, \textit{Open Source}, \textit{CERT}, \textit{Hosted Service}, \textit{Bug Bounty Provider}). We excluded organizational type \textit{consortium} as it has no entries in our dataset. In addition, we incorporate three quantitative factors: the order of CVSS assignment between the CNA and the NVD (i.e., whether the CNA or the NVD published the CVSS vector first), the description length of each CVE, and the CWE. Furthermore, six interaction variables are included to capture whether the effect of being the first to publish a CVSS score differs across organization types. 


\subhead{Organizational Types} To understand the practical impact in the probability of divergent due to a particular factor, we use the following equation reverting the regression equation:
{\small
\begin{eqnarray}
    p_{\overline{X}} & = & Pr\{V_{NVD}\neq V_{CNA}|type(CNA)\neq X\} \\
    p_{X} & = & Pr\{V_{NVD}\neq V_{CNA}|type(CNA)=X\} \\
    \frac{p_X}{1-p_X} &=& \frac{p_{\overline{X}}}{1-p_{\overline{X}}}e^{\beta_X} \label{eq:logistic}
\end{eqnarray}
}
where $e^{\beta_X}$ is the odds ratio associated with CNA type $X$, derived from the logistic regression coefficient $\beta_X$. For example, the logistic (Log.) regression in Table~\ref{tab:RegressionNew} reports a coefficient of +0.653 for the \textit{Vendor}. Exponentiating this coefficient yields an odds ratio of $e^{0.653} \approx 1.9$, indicating that, after controlling for other factors, the odds that a Vendor CNA disagrees with the NVD are approximately 1.9× higher than the odds for non-Vendor CNAs. In other words, \textit{Vendor} CNAs exhibit a substantially higher likelihood of producing CVSS scores that divergent from the NVD, relative to CNAs from other organizational types.

Logistic regression in Table~\ref{tab:RegressionNew} shows that CNA type is significantly associated with the likelihood of rating discrepancies compared to NVD: \textit{Open Source} ($e^{-0.265}\approx 0.8$) and \textit{Hosted Service} ($e^{-1.544}\approx 0.2$) are significantly less likely to produce divergent scores compared to other CNA types, indicating closer alignment with centralized NVD scoring practices.

On the other hand, the odds of divergent with the NVD relative to agreeing with it are higher for \textit{Vendor} (1.9×), \textit{Researchers} (9.8×), and \textit{CERT} (1.8×), compared to all other CNA types. This pattern suggests that these CNA types tend to follow more subjective evaluation practices, leading to greater variability in their scoring outcomes. When \textit{Vendor} and \textit{CERT} disagree with NVD, they tend to assign higher severity scores than the NVD, with a gap of 1.2/10 scores. These values might be enough
to toggle the qualitative class in which the vulnerability is
classified in terms of prioritization. In contrast, \textit{Researchers} tend to assign lower scores than the NVD, with negative coefficients -1.1/10.

\begin{tcolorbox}[
    colback=gray!10,    
    colframe=black,    
    boxrule=0.4pt,     
    arc=1pt,           
    left=4pt, right=4pt, top=2pt, bottom=2pt, 
    enhanced,
    sharp corners
]
\textbf{Takeaway 4.} Structured actors (\textit{Vendor}, \textit{Researcher}, and \textit{CERT}) show higher divergency; decentralized actors (\textit{Open Source}, \textit{Hosted Service}) align more with NVD, reflecting differences in access to internal information, organizational context, and risk modeling practices. 
\end{tcolorbox}

\subhead{First-Mover Effect}
\label{subsec:who first}
We study temporal dependence by determining \emph{Who First} for each CVE. 
When CNA timing is missing in the CVE Program, we compare the earliest CNA appearance in the NVD history to the NVD’s first CVSS date; when CNA timing is missing in the NVD history, we compare the NVD’s first CVSS date to the CNA’s timestamps from the CVE Program. 
If both are present, we take the earlier of (CVE Program CNA update, CNA appearance in NVD) and compare it to the NVD’s first CVSS date. 
We label the outcome as \textit{CNA-First}, \textit{NVD-First}.

To compare all cases, we encode the variable as a binary indicator, assigning 1 to ``CNA First'' and 0 to ``NVD First'', and included it as an independent variable in our regression analysis. As shown in Table~\ref{tab:RegressionNew}, the logistic regression result reveals a positive and statistically significant coefficient (2.072, ***), meaning that the odds of CNAs and NVD disagreement are about 8× higher when the CNA publishes its CVSS score first than when the NVD publishes first.

\begin{tcolorbox}[
    colback=gray!10,    
    colframe=black,    
    boxrule=0.4pt,     
    arc=1pt,           
    left=4pt, right=4pt, top=2pt, bottom=2pt, 
    enhanced,
    sharp corners
]
\textbf{Takeaway 5.} Assessment order significantly affects divergency, with CNA-first assessments generally increasing disagreement. This effect is mitigated for structured actors and amplified for decentralized actors, suggesting differences in information availability and assessment context.
\end{tcolorbox}





\subhead{CWE Effect} We introduce a binary variable, \textit{CWE}, where a value of 0 indicates that both the CNA and NVD assign the same CWE value, while a value of 1 captures all other cases, including mismatched CWE assignments or missing values in either source. Entries labeled as \texttt{NVD-CWE-noinfo} and \texttt{NVD-CWE-Other} are treated as missing values. 
The logistic regression shows a statistically significant positive effect for \textit{CWE}, suggesting that differences in CWE assignments may partly explain, and potentially constitute an underlying source of, CVSS score divergence between CNAs and NVD.

\begin{tcolorbox}[
    colback=gray!10,    
    colframe=black,    
    boxrule=0.4pt,     
    arc=1pt,           
    left=4pt, right=4pt, top=2pt, bottom=2pt, 
    enhanced,
    sharp corners
]
\textbf{Takeaway 6.} Divergency is significantly more likely when CWE assignments are mismatched or missing between CNAs and NVD, suggesting that divergent/missing CWE information may introduce ambiguity into assessment.
\end{tcolorbox}



\subhead{Description Length} 
We also incorporate description word count as a predictor. The results show a statistically significant but practically negligible effect: \(e^{-0.002} \approx 0.998\), indicating an approximately 0.2\% reduction in the odds of divergence per additional word. This aligns with interview results from NVD analysts: when key information is missing, they may need to score vulnerabilities based on a worst-case interpretation. Longer descriptions may therefore reduce divergence by providing additional contextual information.



\subhead{Assessment Order across Organization Types}
While the previous analyses reveal which organization types are more likely to diverge or converge with the NVD, and that overall, when CNAs rate first, the NVD tends to diverge with them, these effects may not be uniform across all organizations. Therefore, we introduce interaction terms between \textit{Who First} and each CNA type (\textit{Who First $\times$ \{Organization Type\}}) to capture potential differences in how early assignment behavior influences divergency across organizations. 

The logistic regression result in Table~\ref{tab:RegressionNew} includes interaction terms between the CNA type and the \textit{Who First} variable (1 = CNA first, 0 = NVD first). These terms capture whether publishing earlier changes the likelihood of a scoring divergence for a given CNA type. A negative interaction coefficient means that, for this CNA type, publishing first reduces the odds of divergence relative to other CNA types. In other words, when the CNA publishes first, the NVD becomes more likely to converge with its assessment. A positive interaction coefficient, in contrast, indicates that publishing first increases the odds of divergence for that CNA type. Equivalently, when the NVD publishes first, the odds of divergence for that CNA type decrease, meaning that its scores tend to be more aligned with the NVD when the NVD leads the scoring process.

\begin{table*}[t]
\caption{Logistic Regression Analysis Results of Each CVSS Metric}
\label{tab:combined-regression-metrics}
\centering
\footnotesize
\resizebox{\textwidth}{!}{ 
\begin{tabular}{l|cc|cc|cc|cc|cc|cc|cc|cc}
\hline
\multirow{2}{*}{\textbf{Factors}}  & \multicolumn{2}{c|}{\textbf{AV Diff}} & \multicolumn{2}{c|}{\textbf{AC Diff}} & \multicolumn{2}{c|}{\textbf{PR Diff}} & \multicolumn{2}{c|}{\textbf{UI Diff}} & \multicolumn{2}{c|}{\textbf{S Diff}}& \multicolumn{2}{c|}{\textbf{C Diff}}& \multicolumn{2}{c|}{\textbf{I Diff}}& \multicolumn{2}{c}{\textbf{A Diff}} \\ \cline{2-17}
 & Coef. & Sig. & Coef. & Sig. & Coef. & Sig. & Coef. & Sig. & Coef. & Sig. & Coef. & Sig. & Coef. & Sig. & Coef. & Sig. \\
\hline

Constant & -2.843 & *** & -3.532 & *** & -1.936 & *** & -3.799 & *** & -2.747 & *** & -2.160 & *** & -2.388 & *** & -2.226 & ***\\ \hline

Vendor & -0.078 &  & 1.431 & *** & -0.208 & *** & 1.278 & *** & 0.470 & *** & -0.629 & *** & -0.184 & ** & -0.094 & \\

Open Source & -0.624 & *** & 0.988 & *** & -0.958 & *** & 0.186 &  & -0.262 & ** & -1.074 & *** & -0.675 & *** & -0.694 & ***\\

Researcher & \textbf{0.820} & *** & \textbf{0.264} & *** & \textbf{1.957} & *** & \textbf{0.396} & *** & \textbf{1.443} & *** & \textbf{2.990} & *** & \textbf{2.242} & *** & \textbf{2.220} & ***\\

Bug Bounty Provider & -0.274 &  & -1.036 &  & -1.386 & * & 0.968 &  & -0.998 &  & 0.695 &  & 0.813 & * & 1.672 & ***\\

Hosted Service & -1.396 & *** & -1.928 & *** & -0.699 & *** & -1.237 & *** & -0.383 & ** & -1.505 & *** & -1.391 & *** & -0.894 & ***\\

CERT & -0.168 &  & 0.684 & ** & -0.344 &  & 0.828 & ** & 0.729 & *** & 0.345 & * & 0.538 &  & 0.471 & ** \\
\hline

Who First & \textbf{1.040} & *** & \textbf{1.152} & *** & \textbf{1.205} & *** & \textbf{0.906} & *** & \textbf{1.410} & ***  & \textbf{1.620} & *** & \textbf{1.504} & *** & \textbf{1.425} & *** \\

CWE & 0.103 & * & 0.022 & & 0.104 & *** & 0.140 & *** & -0.190 & *** & 0.093 & *** & 0.128 & *** & 0.142 & *** \\

Description Length & 0.001 &  & -0.001 & ** & -0.003 & *** & -0.001 & * & 
0.001 & *** & -0.001 &  & -0.001 & * & -0.001 & *** \\


\hline

Who First $\times$ V & -0.716 & *** & -0.703 & *** & -0.450 & *** & -0.211 &  & -0.729 & *** & -0.617 & *** & -0.685 & *** & -0.695 & ***\\

Who First $\times$ OS & 0.466 & ** & -0.270 & ** & 0.877 & *** & -0.572 & *** & 0.634 & *** & 0.822 & *** & 0.642 & *** & 0.640 & ***\\

Who First $\times$ R & -1.614 & *** & -0.531 & *** & -2.068 & *** & 0.158 &  &  -1.466 & *** & -2.223 & *** & -1.755 & *** & -1.701 & ***\\

Who First $\times$ BBP & -0.222 &  & 0.451 &  & 0.778 &  & -2.510 & *** & 0.962 &  & 0.012 &  & 0.221 &  & 0.931 & ** \\
 
Who First $\times$ HS & 0.219 &  & 0.446 & * & -0.267 &  & 0.023 &  & -0.356 & * & 1.113 & *** & 0.787 & *** & 0.532 & **\\

Who First $\times$ CERT & -0.324 &  & -0.846 & ** & -0.656 & ** & -0.080 &  & -0.983 & *** & -1.353 & *** & -1.141 & *** & -1.097 & ***\\
\hline

pseudo-$R^2$ & \multicolumn{2}{c|}{0.053} & \multicolumn{2}{c|}{0.075} & \multicolumn{2}{c|}{0.071} & \multicolumn{2}{c|}{0.068} & \multicolumn{2}{c|}{0.030} & \multicolumn{2}{c|}{0.132} & \multicolumn{2}{c|}{0.074} & \multicolumn{2}{c}{0.144} \\
\hline

\end{tabular}
}
\raggedright 
\scriptsize{
\textbf{Significance levels:}
$^{*} p < 0.05$,
$^{**} p < 0.01$,
$^{***} p < 0.001$.
Non-significant results ($p \ge 0.05$) are left blank.
\textbf{Note:}  Logistic regression on all $43\,997$ entries, dependent variable for each variable is whether there is a difference in scoring for the specific CVSS Metrics. E.g. for the first two columns AV measure the probability that there is disagreement on AV.}
\end{table*}


\textit{Bug Bounty Provider} follows the overall trend that when CNAs rate first, the NVD is more likely to diverge. Both the main effect for \textit{Bug Bounty Provider} and the \textit{Who First} are positive, and their interaction term is also positive (+1.935), indicating a strong amplification effect. Specifically, when \textit{Bug Bounty Provider} publishes first, the likelihood of divergence increases substantially compared to the baseline case. 


For \textit{Vendor}, \textit{Researcher}, and \textit{CERT}, the main logistic effects are positive, meaning these CNA types are generally more prone to diverge with NVD scores. However, their interaction coefficients with \textit{Who First} are negative, meaning that when these CNAs publish their scores before the NVD, the NVD is more likely to converge with them. 
The linear regression gives additional insight; the interaction is not statistically significant for \textit{Vendor} and \textit{CERT}, whereas for \textit{Researcher}, when divergence occurs, they tend to assign scores about 0.8/10 points higher than the NVD.

\textit{Open Source} and \textit{Hosted Service} exhibit negative main effects, indicating greater convergence with the NVD. However, their positive interaction with \textit{Who First} suggests that when they publish first, the likelihood of divergence increases. The linear model further shows that, in cases of divergent, NVD scores tend to be about \(-0.4/10\) points lower than those assigned by \textit{Open Source} and \textit{Hosted Service}.

\subsection{Metric-Specific Divergence}
\label{subsec: metric-specific}

There might be several different dimensions in which divergence materializes, and we now examine how different factors react to individual CVSS metrics. For each CVE, we define a binary dependent variable (e.g., \texttt{AV\_Diff} $=$ 1 if the NVD and CNA assign different values for AV, and 0 otherwise). As shown in Table~\ref{tab:combined-regression-metrics}, the results indicate that the determinants of divergent vary across metrics. The \textit{Who First} variable shows consistently strong and significant effects across all metrics, indicating that publication order is an important factor. In the following, we focus on metric-specific patterns.

\subhead{Impact} \textit{Researchers} show consistently positive and significant coefficients across CVSS metrics, with the strongest effects on \texttt{C} (19.9x), \texttt{I} (9.4x), and \texttt{A} (9.2x). This suggests that divergence is especially likely for \textit{Impact} metrics. The negative \textit{Who First × Researcher} interactions indicate that this effect is weaker when researchers publish first, and more pronounced when NVD publishes first.

Another notable source of divergent comes from \textit{Bug Bounty Providers} on the \textit{Impact} metrics. The logistic regression shows large positive effects for \texttt{I} (2.2x) and \texttt{A} (5.3x). One plausible explanation is incentive-driven reporting: bug bounty platforms aggregate submissions from independent researchers, who may emphasize certain impacts to strengthen reward eligibility. Consequently, reported impact severity may partly reflect bounty incentives rather than a broader consensus view.

\subhead{Attack Complexity} 
\textit{Vendor} and \textit{Open Source},  providers exhibit elevated divergency on the \texttt{AC} metric, with odds ratios of approximately 4.2× and 2.7×, respectively. Confirmed in our interview and implementation, this metric is known to be one of the most difficult to evaluate (also studied for both students and security professionals~\cite{allodi2020measuring}) as it requires a precise understanding of the technical conditions affecting exploitation. It is therefore reasonable to expect that software vendors, who have deeper knowledge of their systems, may diverge with the NVD when it assigns a score first (as reflected by the negative coefficient of ``Who First $\times$ V''). Conversely, when vendors publish their assessments earlier, the NVD may be more inclined to follow.

\subhead{User Interaction} We observe a consistent pattern of positive coefficients for the \texttt{UI} metric across multiple CNA types, with statistically significant effects for \textit{Vendor}, \textit{Researcher}, and \textit{CERT}. Despite appearing conceptually simple \cite{allodi2020measuring}, the interpretation of UI (e.g., whether a victim must ``click'' or otherwise perform an action) remains ambiguous in several corner cases such as Cross-Site Scripting (XSS). This ambiguity has been widely discussed within the CVSS SIG, they dedicated several meetings after CVSS v4 was released to clarify these aspects and revised its user guide (\url{https://www.first.org/cvss/v4-0/faq}).

\begin{tcolorbox}[
    colback=gray!10,    
    colframe=black,    
    boxrule=0.4pt,     
    arc=1pt,           
    left=4pt, right=4pt, top=2pt, bottom=2pt, 
    enhanced,
    sharp corners
]
\textbf{Takeaway 7.} Divergency is significantly associated with \texttt{AC}, \texttt{UI}, and \texttt{Impact} metrics, with some CNAs showing stronger divergence in these dimensions. These metrics are more ambiguous, aligning with refinements in CVSSv4 that further distinguish \texttt{AC}, \texttt{UI}, and \texttt{Impact} metrics.
\end{tcolorbox}




\section{Root Causes Analysis} \label{subsec:humaninterview}

Based on feedback from the NVD, six CNAs, and seven additional experts who participated in the subsequent CVSS SIG discussion, we identified eight recurring root causes (RCs) of divergence between CNAs and the NVD. We discuss each RC below and provide 12 representative examples in Table~\ref{tab:root_causes}.

\subhead{RC1 - Divergence is not an error} Divergence is not an error but actually a right thing. Some assessments are based on downstream product configurations and deployment contexts, which may differ from upstream assumptions. For example, security features, such as SELinux, may be enabled by default in downstream distributions, reducing the exploitability of certain vulnerabilities. As a result, CNAs may assign lower severity scores, as they adopt a more ``secure'' default stance compared to some upstream repositories.

\subhead{RC2 - Differences in CWE/CPE} One CNA mentioned that vulnerabilities with identical descriptions may still be assigned different CWE values, and that such differences should be considered when analyzing divergence. They also noted the importance of verifying whether other attributes, such as CPE (i.e., the affected products), are also the same. This insight suggests that, at the description level, relying solely on textual similarity may be insufficient; additional structured information, such as CWE and CPE, should also be taken into account (See the detailed results in Table~\ref{tab:cna_or_entropy}).

In addition, feedback from the NVD indicates a potential cause of CWE discrepancies. CWE is maintained by MITRE, and the NVD uses only a subset of the full CWE taxonomy for enrichment. As a result, the NVD may not always have access to the same level of specificity as the CWE values provided by CNAs, leading to differences in assigned CWE categories.

\subhead{RC3 - Selective Disclosure of Scoring Rationale}
The CVSS scores from some CNAs often align with the NVD because additional scoring rationale is provided in the frequently asked questions (FAQs) accompanying their advisories. These FAQs offer further explanations for metric assignments (e.g., AC:H or S:C), enabling NVD analysts to assign the same CVSS vector from publicly available information. However, the internal variability remains substantial, and they typically tend to provide explanations only when metric values deviate from commonly expected settings (e.g., explaining AC:H but not AC:L). Therefore, even when additional explanations are available, divergences may persist due to the absence of a complete and systematic description of the scoring rationale.


\subhead{RC4 - Human Errors} Divergency may also arise from CVE submission process. For example, vulnerability descriptions in the NVD may not match those provided in the original security advisories. One CNA indicated that such divergence can occur when incorrect or incomplete data is submitted to the CVE system, causing NVD to base its assessment on inaccurate information. 


\subhead{RC5 - Insufficient Detail in CVE Descriptions} Inconsistencies in CVSS scoring may stem from insufficient detail in CVE descriptions. In some cases, descriptions do not provide enough information to justify specific metric assignments, making it difficult for external analysts to reach the same conclusions (e.g., NVD). For example, one CNA noted that descriptions may lack sufficient detail to explain why certain metrics (e.g., PR) are rated as Low versus High. Another CNA further suggested that CVE descriptions may not accurately convey all aspects of the vector. Encouragingly, one CNA reported that more detailed descriptions have started to appear in executive summaries after 2025, providing additional insight into the vulnerability. We conducted an extra study after 2025 in Section~\ref{sec:2025}.

In other cases, similar or identical descriptions may be used for vulnerabilities that differ in their actual impact. For instance, two XSS vulnerabilities with the same description may exhibit different impacts depending on the exploitation context. One CNA suggested that the root cause may lie in the fact that CVE descriptions are not sufficiently curated to distinguish each individual XSS vulnerabilities, particularly in terms of their impact differences.


\subhead{RC6 - Information Asymmetry and Limited Context for NVD} Discrepancies may be caused by information asymmetry, where CNAs possess internal information that is not accessible to the NVD. For example, the information available to the NVD may be incomplete or inconsistent with that held by CNAs. Additionally, the NVD may lack access to proof-of-concept exploits or sufficient contextual data required to consistently assess CVSS scores with high accuracy. One CNA further suggested that, in practice, it may be more reliable to rely on CVSS scores assigned by CNAs, as the NVD must process a large volume of CVEs annually, making it impractical to perform deep technical validation for each entry.

\subhead{RC7 - Risk Modeling Differences} Some discrepancies arise from differences in how organizations model and interpret risk. This divergence in risk modeling assumptions can lead to systematic differences in CVSS scoring across sources. Some discrepancies arise from vulnerabilities that involve multiple distinct attack scenarios. In such cases, a single CVE may encompass different exploitation paths with varying levels of impact and attack complexity. For example, one CNA described a vulnerability that includes both a race condition requiring high attack complexity (AC:H) and a simpler attack path with low attack complexity (AC:L) leading to availability impact. The CNA chose to assign the score based on the highest impact scenario. However, other CNAs may combine all aspects of the vulnerability description into its assessment, even when these correspond to distinct attack scenarios.

\subhead{RC8 - Ambiguity in CVSS Specification} One CNA suggested that inconsistencies may arise from inherent ambiguity in the CVSS specification itself.  This is supported by the evolution of the CVSS standard, where several refinements have been introduced in CVSS v4 to address potential ambiguities in earlier versions. For example, CVSS v4 separates Attack Complexity in CVSS v3.1 into two distinct metrics: Attack Complexity and Attack Requirements. 

\begin{tcolorbox}[
    colback=gray!10,    
    colframe=black,    
    boxrule=0.4pt,     
    arc=1pt,           
    left=4pt, right=4pt, top=2pt, bottom=2pt, 
    enhanced,
    sharp corners
]
\textbf{Takeaway 8.} As shown in Table~\ref{tab:root_causes}, we observe that many examples involve insufficient detail to support metric assessment and information asymmetry across sources. Divergent assessments may therefore be both correct state unless more information is provided in individual CVE.
\end{tcolorbox}


\begin{table}[t]
\centering
\scriptsize
\setlength{\tabcolsep}{3pt}
\caption{CVE Examples of Root Causes}
\label{tab:root_causes}
\begin{tabular}{c|c|c|c|c|c|c|c|c|c}
\hline
\# & \textbf{CVE-ID} & \textbf{RC1} & \textbf{RC2} & \textbf{RC3} & \textbf{RC4} & \textbf{RC5} & \textbf{RC6} & \textbf{RC7} & \textbf{RC8} \\
\hline
1 & CVE-2024-28907 & & x & x &  &  &  & & \\\hline
2& CVE-2024-30007 & & x & x &  &  &  &  &\\\hline
3&CVE-2021-1471  & &  &  & x &  & x &  &\\\hline
4&CVE-2022-20955 & &  &  & x &  & x &  &\\\hline
5&CVE-2020-3393  & &  &  &  &  & x &  &\\\hline
6&CVE-2025-20270 & &  &  &  &  & x & & \\\hline
7&CVE-2024-20509 & &  &  &  &  &  & x &\\\hline
8&CVE-2020-1227  & &  &  &  & x &  &  &\\\hline
9&CVE-2020-16945 & &  &  &  & x &  &  &\\\hline
10&CVE-2022-21830 & &  &  &  & x &  & & \\\hline
11& CVE-2022-0847 & &  &  &  & x &  &  &\\\hline
12& CVE-2019-5736 & x &  &  &  & x & x & & x \\\hline
\multicolumn{2}{c|}{Total} & 1 & 2 & 2 & 2 & 5 & 5 & 1 &1\\ \hline
\end{tabular}
\end{table}



\section{Case Studies}
\label{subsec:case-study}
To better understand the root causes in practice, we conduct a case study with two analysts from a CNA who have implemented real-world attacks for two vulnerabilities: Dirty Pipe (CVE-2022-0847) and runC Overwriting (CVE-2019-5736), corresponding to the last two CVEs in Table~\ref{tab:root_causes}. Both attack implementations are conducted on Ubuntu 22.04.

For each case, we first introduce the CVSS metric definitions to the analysts. Based on their implementation experience, we ask them to assign values for each CVSS metric and assess whether the corresponding CVE description provides sufficient information to support these assignments. Finally, we discuss root causes for other cases. 

\subhead{Dirty Pipe} 
This vulnerability is assigned identical CVSS metrics by the NVD, \textit{SUSE} and \textit{CISA-ADP}. Based on the implementation experience of the analysts, they assigned the vector \texttt{CVSS:3.1/AV:L/AC:L/PR:L/UI:N/S:U/C:H/I:H/\\A:H}, which matches the CVSS vector provided by these three sources. However, they indicated that the information related to \texttt{AC} is insufficient and that \texttt{UI} is unclear (RC5).

\subhead{runC Overwriting} 
This vulnerability is assigned different CVSS vectors by the NVD and multiple CNAs, including \textit{SUSE}, \textit{RedHat}, and \textit{Canonical}. Notably, \textit{SUSE} and \textit{RedHat} share the same vulnerability description, while \textit{Canonical} shares the same description as the NVD (RC6). As shown in Table~\ref{tab:cve5736}, disagreement is primarily concentrated in the \texttt{AC} and \texttt{PR} metrics. Sources with identical descriptions tend to assign consistent values for \texttt{AC}, \textit{SUSE} and \textit{RedHat} assign \texttt{AC:H}, whereas the NVD and \textit{Canonical} assign \texttt{AC:L}, suggesting that vulnerability descriptions strongly influence metric assignment. 

However, discrepancies remain even under identical descriptions. Major sources assign \texttt{PR:N}, whereas the analysts assign \texttt{PR:H} based on their implementation experience, explaining that root privilege inside the container is required to perform the attack. This indicates that existing sources may value differently the privilege requirement. The analysts assign \texttt{AC:L}, consistent with the NVD and \textit{Canonical}, likely because their implementation was conducted on Ubuntu, a Linux distribution maintained by \textit{Canonical} (RC1).

When assessing description quality, the analysts report that the \textit{SUSE} and \textit{RedHat} descriptions lack sufficient information for \texttt{AC}, \texttt{PR}, and \texttt{UI}, while the NVD and \textit{Canonical} descriptions provide slightly more information for \texttt{AC}, and \texttt{UI} remains unclear (RC5).

These two cases highlight the importance of providing accurate and detailed descriptions. Divergency may arise from differences in descriptions across sources or missing information for certain metrics, especially \texttt{AC}, which the analysts also identify as the most difficult to assign, as it depends on the interpretation of attack complexity (RC8) and involves many subjective judgment.

\subhead{Cases with RC2\&RC3} CVE-2024-28907 and CVE-2024-30007 contain identical description but different CVSS scores. Augmenting with their CWEs (CWE-59 and CWE-269), both CVEs should fall into different groups, thus resolving the divergence.


\subhead{Cases with RC4\&RC6} CVE-2021-1471 and CVE-2022-20955 illustrate discrepancies arising from the CVE submission process. Incorrect/incomplete data may be submitted to the CVE system, leading to mismatches between NVD descriptions and original security advisories, which can cause disagreement across sources.

\subhead{Cases with RC7} 
CVE-2024-20509 involves multiple exploitation paths, including a race condition (\texttt{AC:H}) and a low-complexity path (\texttt{AC:L}). Different sources may prioritize different paths, resulting in divergent metric assignments.
 
\begin{table}[t]
\centering
\scriptsize
\caption{CVSS across different sources (CVE-2019-5736)}
\label{tab:cve5736}
\begin{tabular}{l l l}
\hline
\textbf{Source} & \textbf{CVSS Vector} \\
\hline

\textit{SUSE} & \texttt{CVSS:3.0/AV:L/\textcolor{red}{AC:H}/\textcolor{red}{PR:L}/UI:R/S:C/C:H/I:H/A:H}\\
\textit{RedHat} & \texttt{CVSS:3.1/AV:L/\textcolor{red}{AC:H}/\textcolor{red}{PR:N}/UI:R/S:C/C:H/I:H/A:H}\\\hline
NVD & \texttt{CVSS:3.1/AV:L/\textcolor{red}{AC:L}/\textcolor{red}{PR:N}/UI:R/S:C/C:H/I:H/A:H}\\
\textit{Canonical} & \texttt{CVSS:3.1/AV:L/\textcolor{red}{AC:L}/\textcolor{red}{PR:N}/UI:R/S:C/C:H/I:H/A:H}\\\hline
\rowcolor{gray!15}
Implemented & \texttt{CVSS:3.1/AV:L/\textcolor{red}{AC:L}/\textcolor{red}{PR:H}/UI:R/S:C/C:H/I:H/A:H}\\

\hline
\end{tabular}
\end{table}

\section{Temporal Analysis and Post-2025 Trends} \label{sec:2025}
As highlighted in RC5, a CNA noted that they have started to provide more detailed descriptions after 2025. We therefore conduct a follow-up analysis using data after January 2025. While our original dataset covers CVEs up to January 2025, we collect additional data from February 2025 to March 2026 to examine whether these divergence issues are reduced over time.

\begin{figure}[t]
\centering
\includegraphics[width=0.6\linewidth]{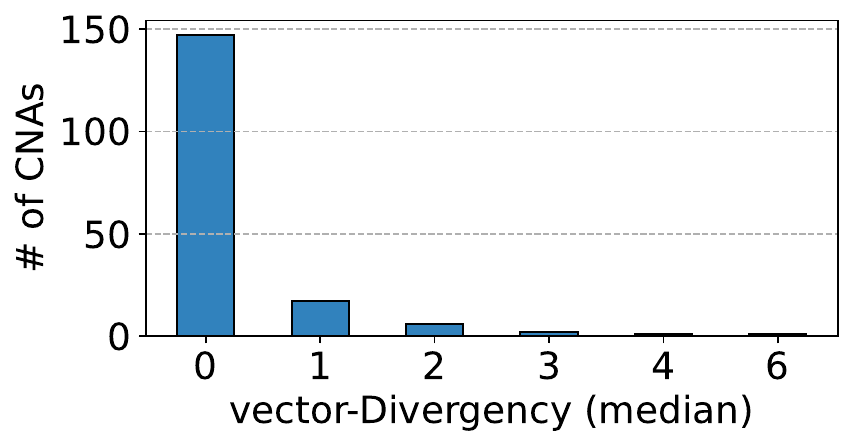}
\caption{\inconmeasure{}-Divergency values under the Consumer-View settings after 2025}
\label{fig:overall-after2025}
\end{figure}

\subhead{\inconmeasure{}-Divergency Analysis After 2025}
Fig.~\ref{fig:overall-after2025} shows how the median \inconmeasure{}-divergency of publicly available CNAs changes after 2025. The results indicate that only \textbf{27/174 (16\%)} of CNAs have a \emph{median} \inconmeasure{}-Disagreement of at least \(1\), representing a substantial reduction in CNA–NVD divergency compared to the pre-2025 result (\textbf{48\%}). This suggests a notable shift toward greater convergency in CVSS scoring after 2025.

\begin{figure}
\centering
\includegraphics[width=\linewidth]{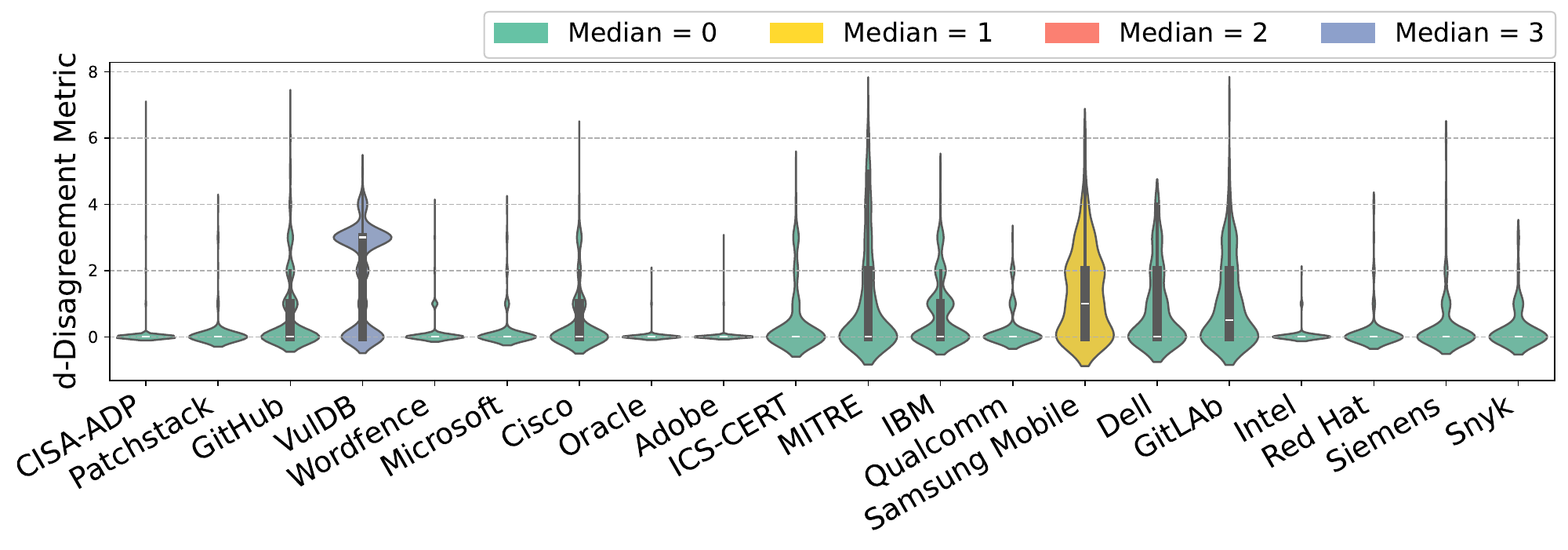}
\caption{Distribution of \inconmeasure{}-level divergency of Top 20 CNAs under Consumer-View after 2025}
\label{fig:violin-after2025}
\end{figure}

Fig.~\ref{fig:violin-after2025} shows the distribution of \inconmeasure{}-divergence values for the top 20 CNAs after 2025 under the \textit{Consumer-View} setting. Notably, only two CNAs exhibit a median \inconmeasure{} of at least \(1\), while the majority are concentrated at 0. Compared to the pre-2025 (Figure~\ref{fig:violin_plot}), most CNAs show reduced divergency. \textit{Samsung Mobile} dropped from median value (\inconmeasure{}\(=2\)) before 2025 to \(1\) after 2025. Only \textit{VulDB} shows no noticeable improvement, continuing to exhibit relatively high divergency (\inconmeasure{}\(=3\)) both before and after 2025.

Overall, these results suggest that the reduction in disagreement after 2025 is widespread across major CNAs, with only a small number of sources still exhibiting persistent divergence.

\begin{figure}
\centering
\includegraphics[width=1\linewidth]{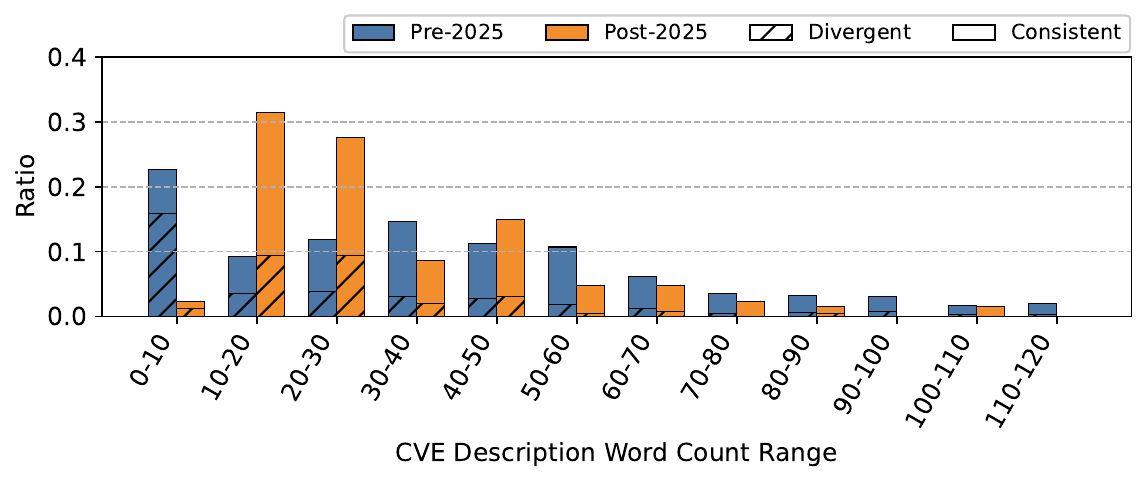}
\caption{Relationship between CVE description length and Description-level divergency before and after 2025}
\label{fig:words_length}
\end{figure}

\subhead{Description-level Divergency After 2025} 
Figure~\ref{fig:words_length} shows a clear shift in description length before and after 2025. The share of short descriptions, i.e., those with 0-10 words, decreased substantially after 2025, from over 20\% to 2.4\%. Meanwhile, descriptions with 20-30 words become the majority group, replacing the 0-10 word group that dominated before 2025. Although the absolute divergence rates in the 10--20 and 20--30 word groups are higher after 2025 than before 2025, these groups account for only about 1/3 of all cases, compared with approximately 50\% before 2025. This provides further evidence that longer descriptions are associated with lower divergence and RC5, i.e.,  more detailed descriptions have started to appear after 2025.

\begin{tcolorbox}[
    colback=gray!10,    
    colframe=black,    
    boxrule=0.4pt,     
    arc=1pt,           
    left=4pt, right=4pt, top=2pt, bottom=2pt, 
    enhanced,
    sharp corners
]
\textbf{Takeaway 9.} The divergency between CNA and NVD decreases after 2025 compared to the pre-2025 period under the \textit{Consumer-View} setting. This indicates improved consistency in CVSS assessments over time, possible due to better description quality.
\end{tcolorbox}



\section{Impact} \label{sec:discussion}
In this section, we discuss  potential impacts on both research and downstream applications.

\subhead{Broad Impact on Research}
To illustrate the impact of the reliability of CVSS scores, we systematically collected papers published between 2021 and 2025 from the A$^\ast$ conferences in security and software engineering, as ranked in the 2023 CORE list~\cite{core2023} (four security conferences: IEEE S\&P, USENIX Security, ACM CCS, and ISOC NDSS, three software engineering conferences: ASE, ICSE, and ESEC/FSE). The impact maybe larger as we have not considered journals (e.g., TSE) or other internet conferences such as IMC or WWW. We searched all papers having CVSS in the text and belonging to these venues from year 2021 to 2025 \footnote{Google advanced search feature with the string ``CVSS'' source: ``Symposium on Security and Privacy'' OR source: ``USENIX Security'' OR source: ``Conference on Computer and Communications Security'' OR source: ``Network and Distributed System Security'' OR source: ``Automated Software Engineering'' OR source: ``International Conference on Software Engineering'' OR source: ``Foundations of Software Engineering''}

This search initially yielded 109 papers. We excluded nine papers that were published in conferences outside the selected seven conferences, despite having similar names, such as Euro S\&P and ASIA CCS, and one USENIX paper~\cite{DBLP:conf/uss/Lin0023} that included the term ``CVSs'' which was unrelated to our context. We then read the full text of the remaining 100 papers. A paper was considered impacted if divergence in CVSS scoring could substantially influence its reported effectiveness, thereby affecting the interpretability, comparability, or external validity of its results. Based on this criterion, we identified 25 impacted papers, which are categorized in Table~\ref{tab:impacted-works-categorized} according to the four impact categories (I1–I4).




\begin{table*}
\caption{Sample of Works in A* Security and Software Engineering Conferences 2021-25 and Why They Are Affected}
\small
\begin{tabular}{p{0.21\linewidth}|p{0.74\linewidth}}
\textbf{Prior Works} & \textbf{Why Impacted?} \\
\hline
\cite{DBLP:conf/uss/WooHCL22}, 
\cite{DBLP:conf/sigsoft/HuWHWKT21}, 
\cite{DBLP:conf/sp/LiuGZGZWG24}, 
\cite{DBLP:conf/sp/KimSSJS24},  
\cite{DBLP:conf/uss/LiJCLLCLWBCL021}, \cite{DBLP:conf/sp/XiaoSYL24},
\cite{DBLP:conf/ccs/TanZMCSLY21}, 
\cite{DBLP:conf/ccs/QinXL23}, \cite{DBLP:conf/kbse/WangLPGCWGL23}, \cite{DBLP:conf/sigsoft/Cheng0BW22}  & 
I1: Use CVSS as input feature or guideline (e.g., filter low CVSS entries), program input point may differ based on sources\\
\hline

 \cite{DBLP:conf/icse/PanBZ00L24}, \cite{DBLP:journals/ase/WangCPY25}, \cite{DBLP:conf/kbse/LeHCB21},  \cite{DBLP:conf/uss/CerdeiraM0022}, \cite{DBLP:conf/uss/JahanshahiANE23}&  
I2: Use CVSS scores to represent vulnerability severity or support decision-making (e.g., patching, mitigation, prioritization), different sources provide different priorities  
\\

\hline
\cite{DBLP:conf/sigsoft/LuoSZWSCSLHJ24},  \cite{DBLP:conf/uss/SuciuNLBD22},
\cite{DBLP:conf/uss/EthembabaogluWZ24}
& I3: Use description to predict CVSS metrics, same description may appear in both training and testing with different scores \\

\hline
\cite{DBLP:conf/sp/WunderKEGB24},
\cite{DBLP:conf/ccs/JiangGHTRTEZP22}, 
\cite{DBLP:conf/sp/WangCXWGWDLW24}, 
\cite{DBLP:conf/sp/PerezEG24}, \cite{DBLP:conf/ndss/Yu0LLLLDWSBH24}, \cite{DBLP:conf/ndss/0007KSJXL25}, \cite{DBLP:conf/uss/KondrackiN24}
 & I4: Use CVSS as evaluation ground truth, different evaluation sources provide different ground truth 
 \\
\hline

\end{tabular}
\label{tab:impacted-works-categorized}
\end{table*}

\begin{figure}[t]
\centering
\includegraphics[width = \linewidth]{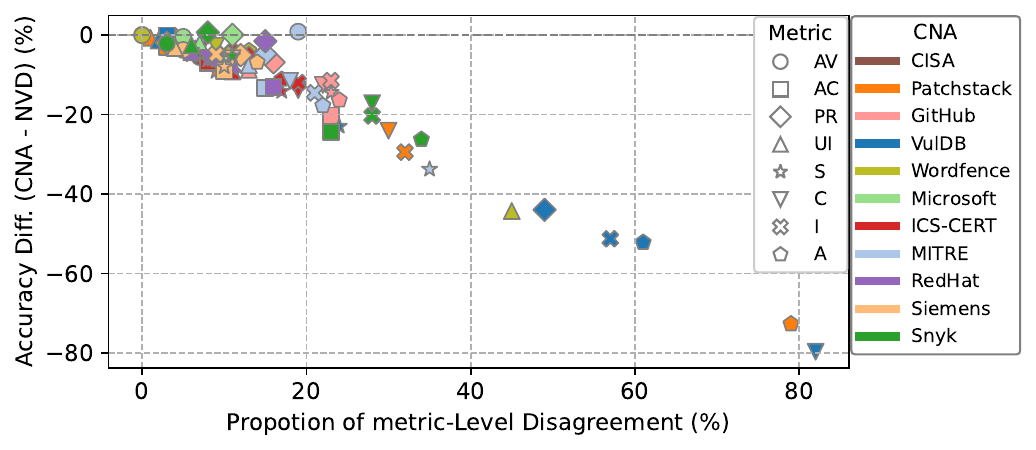}
\caption{The same CVE rated by a CNA and NVD}
\label{fig:downstreamImpact}
\end{figure}

\subhead{Downstream Impact} We reproduced the CVSS-BERT model of Shahid et al.~\cite{DBLP:conf/icmla/ShahidD21}, which predicts CVSS base metrics from descriptions. 
We followed the original training setup and evaluated \emph{per CNA}, removing all overlapping (identical) descriptions from training to avoid leakage. As shown in Fig.~\ref{fig:downstreamImpact}, accuracy \emph{systematically drops} across CNAs. In particular, for \textit{Wordfence}, where \texttt{UI} disagreements are amplified by 957 entries, \texttt{UI} accuracy falls by \(\sim\!40\%\) relative to the reported baseline. 
This is not a single case: \textit{VulDB} shows similarly large degradations on \texttt{PR} and the \texttt{Impact} metrics.

These results indicate that, without explicitly quantifying NVD and CNA inconsistencies, models trained on one source do not reliably generalize to others. Consequently, reported metrics depend heavily on the chosen test source, resulting non-comparable scores. This motivates source-aware discrepancy metrics (external distance and internal stability) and per-source evaluation.

\subhead{Implications for CVSS Framework}
For CVSS practice, these disagreements reveal where
additional guidelines may be necessary (e.g., longer description, error-prone vectors, etc.), and the ongoing development of CVSSv4 can help simplify rating dimensions such as Attack Complexity. Our proposed metrics further support continuous evaluation of cross-version consistency when
more CVSSv4 data become available.




\subhead{Implication for Practice}
Organizations should expect and plan for systematic divergence between CNA and NVD assessments, even among closely related CNAs. Lacking vulnerability information and different interpretation of ``worst case''~\cite{nvd-worsecase} can explain much of this disagreement. Relying on a single source introduces bias; integrating multiple sources without clear rules can propagate inconsistencies. We therefore recommend explicit, source-aware selection and tie-breaking policies when integrating CVSS data from multiple sources.

Even within a single source, internal assessments may conflict: CVEs with \emph{identical descriptions} and matching CWE and CPE attributes can still receive different vectors. Such inconsistencies stem from manual errors~\cite{DBLP:conf/cloudcom/ZhangCZZC23} and subjective interpretations~\cite{DBLP:journals/tnn/FrenayV14}. This can mislead downstream consumers and underscores the need for stricter CNA hygiene, including  more informative descriptions.

\subhead{Compliance and governance} The assessment discrepancies may impact the effectiveness of vulnerability management practices and undermine compliance with standards and laws~\cite{iso27005,nist80053,pci_dss,cra} that require organizations to assess and prioritize vulnerabilities based on reliable criteria, and that mandate timely remediation of high-risk vulnerabilities. Source-aware integration and stability filters (OR) improve auditability and help ensure consistent, defensible prioritization.

\subhead{Implications for Research} Our study impacts several areas of software security research that rely heavily on CVSS. The most apparent impact is that when preparing datasets for machine learning models, filtering and data cleaning is necessary to reduce label noise and improve model reliability, even from trusted sources. As summarized in Table~\ref{tab:impacted-works-categorized}, prior works on vulnerability severity prediction, patch prioritization, and vulnerability analysis directly depend on NVD CVSS data may inherit biases or noise from the underlying datasets. The effectiveness of vulnerability mining works may be substantially influenced by changing rating methods. Therefore, before using CVSS data for downstream security processes, researchers should carefully address discrepancies both within and across sources to improve model reliability and accuracy.

We already observed length of the description contributes to divergency in the regression results. For downstream vulnerability modeling, automated scoring, and user studies, descriptions that are extremely short should be handled with care: 
either \textbf{filter} short descriptions or \textbf{augment} them with contextual details (e.g.,CWE,  affected CPEs, impact scope, exploit prerequisites) to improve model robustness and reduce ambiguity in human evaluations.

\section{Limitations}
\label{sec:limitations}

\subhead{Imbalanced Data Entries from CNAs} To reduce selection bias, we include all CNAs that have published CVSS scores in the NVD, rather than focusing solely on a few prominent organizations. This broad coverage improves the internal validity of our findings, ensuring that our analysis reflects the full spectrum of CVSS assessment behaviors in the ecosystem. However, the reliability of our consistency measurements, such as odds ratio  is inherently influenced by the amount of available data per CNA. For some CNAs with limited data or few entries sharing identical descriptions, our metrics may lack sufficient statistical power to support strong conclusions. In contrast, for well-represented CNAs with many comparable groups (e.g., Oracle), consistent scoring behavior is supported by both high data volume and low disagreement scores.

\subhead{NVD Status and Consumer Assumption} 
Our analysis does not explicitly consider the NVD processing status (e.g., whether a CVE is still under modification or fully analyzed). In practice, NVD recommends considering such status information when interpreting CVSS scores. However, from a consumer perspective, users typically rely on available scores without inspecting processing status. Therefore, we approximate a \textit{Consumer-View} setting by treating CNA-only entries as implicitly followed by NVD. While this assumption reflects practical usage, it may overestimate agreement when NVD scores are still evolving. Incorporating finer-grained status information is an important direction for future work.


\subhead{Generality to CVSS v4.0} 
Although our study is based on CVSS v3.1, we believe that our methodology is still applicable under CVSS v4.0, despite structural changes and added metrics in the new version. The approach remains capable of identifying discrepancies in severity assessments across sources, offering potential value for both industrial systems and academic research that rely heavily on consistent vulnerability evaluation.

\section{Related Work}
\label{sec:related}

\subhead{Features Extraction for CVSS} Several research efforts extract specific features, such as vulnerability descriptions and assessment information from public repositories like the NVD~\cite{DBLP:conf/fps/WeerawardhanaMR14,DBLP:conf/uss/DongGCXZ019,DBLP:conf/dbsec/ZhangZZ23,DBLP:conf/kdd/BinyaminiBIYES21,ni2022predicting}, which can be used to predict CVSS base metrics values~\cite{DBLP:conf/iceccs/GongXLFH19,DBLP:conf/badgers/YamamotoMN15,DBLP:journals/access/CostaRSPI22,DBLP:conf/IEEEares/ElbazRM20}. Studies~\cite{DBLP:conf/ijcai/ChenLLPS19,DBLP:journals/corr/abs-1801-00938} note that manual severity assessments often take over 130 days to complete, prompting the development of ML-based automation approaches. Early work relies on traditional models like SVM~\cite{DBLP:conf/scai/EdkrantzS15} and XGBoost~\cite{manai2024helping}, while later studies shift toward neural networks and NLP-based models, such as BERT-based classifiers~\cite{DBLP:journals/kbs/YinTCW20,DBLP:conf/icmla/ShahidD21,DBLP:journals/access/CostaRSPI22,ni2022predicting}. More recently, large language models (LLMs) are explored for CVSS classification and cybersecurity entity alignment across multiple data sources~\cite{marchiori2025llmsclassifycvesinvestigating,DBLP:conf/cloudnet/MirandaSMSKRLL24,DBLP:conf/ccs/QinXL23}.

However, these works assumes CVSS annotations across public sources are consistent and reliable, without mitigation. Our study systematically investigates the prevalence and impact of scoring divergencies, providing insights for building more trustworthy ML-based vulnerability analysis.

\subhead{Disagreement/Inconsistency in Vulnerability Databases} Multiple works study the inconsistency in vulnerability data sources, such as severity scores and vulnerability types~\cite{DBLP:journals/tdsc/AnwarACLM22}, mismatched software names and versions~\cite{DBLP:conf/uss/DongGCXZ019},  CVSS base metrics assigned by two different organizations~\cite{DBLP:journals/array/JiangA22}, and discrepancies between same or semantically similar NVD entries~\cite{DBLP:conf/cloudcom/ZhangCZZC23}. A user study by Wunder et al.~\cite{DBLP:conf/sp/WunderKEGB24} survey 196 CVSS users, revealing a high degree of subjectivity in severity assessments. Mell et al.~\cite{mell2022cvss} investigate the gap between CVSS scoring and expert opinion. A recent study further shows the difference in scoring between different vulnerability scoring systems, leading to conflicting prioritization signals~\cite{DBLP:conf/ccs/KoscinskiNOFM25}.


While some prior work has noted divergence in vulnerability databases, it lacks a systematic analysis of such divergence and the reliability of original CNAs. To fill this gap, we analyze how divergence correlates with CNA types and introduce odds ratio to identify more consistent data subsets.

\subhead{CNA-Based Rating} Beyond individual-level variations, institutional inconsistencies are also observed. Coutinho et al.~\cite{DBLP:conf/ladc/CoutinhoM0LKRKL24} analyze how contextual, product-specific factors influence CVSS score divergence, but their in-depth analysis is limited to ten CNAs, leading to misleading takeaways such as ``NVD generally assigns higher CVSS scores than CNAs''. In contrast, our regression across all CNAs reveals that scoring tendencies vary by organizational type. For example, vendors tend to assign higher scores than the NVD, whereas research-oriented CNAs may often assign lower ones (see Table~\ref{tab:RegressionNew}). Similarly, Miranda et al.~\cite{DBLP:conf/cloudnet/MirandaSMSKRLL24} employ large language models (LLMs) to learn CNA-specific CVSS scoring patterns, achieving over 70\% accuracy and F1-scores, suggesting the potential for knowledge transfer across CNAs in public datasets, e.g., NVD.

While these studies focus primarily on product-specific vulnerabilities and investigate possible knowledge transfer between CNAs and the NVD, our work provides a broader perspective.
We systematically analyze CNA–NVD discrepancies across all available CNAs to examine overall assessment orientations.
Rather than focusing on discussing a few individual organizations, our regression approach uncovers how organizational types and other contextual factors shape divergence patterns in CVSS scoring.

\section{Conclusion} \label{sec:con}

In this paper, we developed systematic metrics to quantify divergency between CNAs and the NVD and introduced the first group-based framework to assess internal divergency by grouping CVEs with identical descriptions. Across all 288 CNAs, 73\% (pairwise) and 48\% (consumer-view) showed at least one metric-level divergency in the majority of their rated CVEs. A statistically significant analysis across the full set of CNAs shows that risk-sensitive organizations, such as \textit{Vendors}, \textit{Researchers} and \textit{CERT}, diverge more frequently. Specifically, \textit{Vendors} and \textit{CERT} often assign higher severity scores, whereas \textit{Researchers} tend to assign lower scores. When CNA releases the CVSS first, NVD tend to disagree with it. However, certain types of CNA, such as \textit{Vendors}, \textit{Researchers}, and \textit{CERT}, tend to trusted by NVD. Consistent across different studies in the paper, the most frequent sources of disagreement are the \textbf{Attack Complexity}, \textbf{User Interaction}, and \textbf{Impact metrics}. Through feedback from CNAs, NVD, and the CVSS SIG, we further summarize eight root causes of divergence and analyze 12 representative vulnerabilities. These cases suggest that insufficient detail in CVE descriptions and information asymmetry are among the most common causes of divergent scoring. Finally, our extended analysis shows a positive trend after 2025, suggesting that more detailed CVE descriptions may help reduce CVSS divergence. The possible impact discussed in this paper sheds light on future directions for researchers, government policymakers, and practitioners. We hope these findings raise awareness among CVE consumers, inform future CVE reporting practices and CVSS guidance, and help the ecosystem move toward more transparent, consistent, and actionable vulnerability assessment.



\bibliographystyle{IEEEtran}
\bibliography{Reference}

\clearpage

\appendix 
\label{sec:appendix}

\begin{figure}[h]
\centering
\footnotesize
\includegraphics[width=\linewidth]{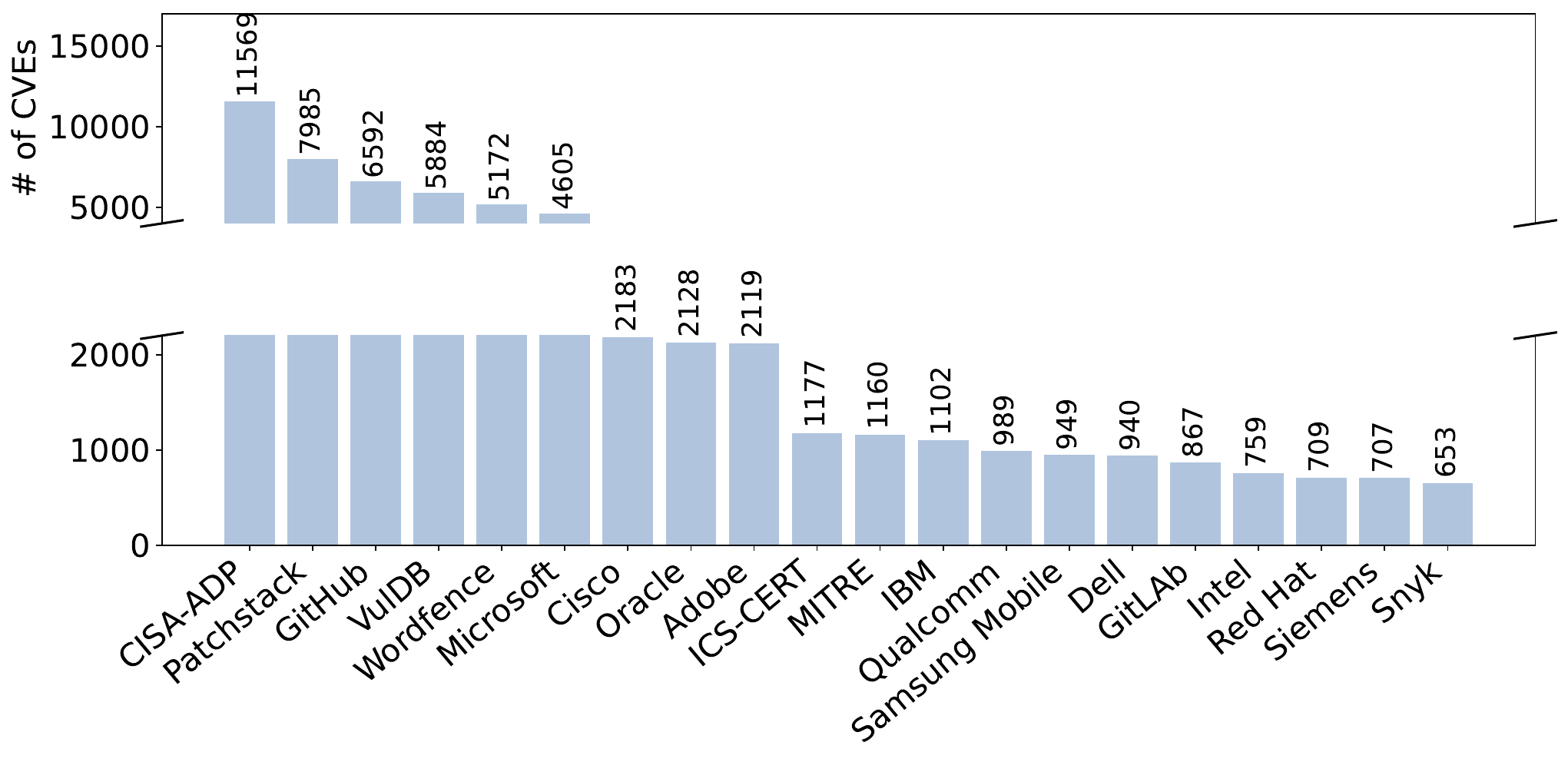}
\caption{\#CVEs Assigned by Top Public 20 CNAs}
\label{fig:CNAs}
\end{figure}

\begin{table}[h]
\centering
\scriptsize
\setlength{\tabcolsep}{3.5pt} 
\caption{Metric-Level Divergent Counts per CNA}
\label{tab:metric_level_disagreement}

\begin{tabular}{l|c|c|c|c|c|c|c|c|c}
\hline
\multirow{2}{*}{\textbf{CNAs}} 
& \multicolumn{8}{c|}{\textbf{Divergency metric\# (metric-Level)}} 
& \multirow{2}{*}{\textbf{\#CVE}} \\ \cline{2-9}
 & AV & AC & PR & UI & S & C & I & A &  \\ \hline

CISA-ADP        & 345 & 169 & 538 & 287 & 136 & 362 & 305 & 356 & 4860 \\ \hline
Patchstack      & 0   & 114 & 153 & 40  & 341 & 1091 & 1166 & 773 & 3601 \\ \hline
GitHub          & 239 & 1234 & 862 & 717 & 1275 & 1211 & 1225 & 1322 & 5441 \\ \hline
VulDB           & 494 & 121 & 2309 & 106 & 1116 & 3896 & 2709 & 2870 & 4742 \\ \hline
Wordfence       & 2   & 161 & 62  & 957 & 130 & 193 & 204 & 282 & 2132 \\ \hline
Microsoft       & 96  & 183 & 205 & 129 & 111 & 171 & 177 & 198 & 1905 \\ \hline
Cisco           & 42  & 102 & 120 & 81  & 228 & 232 & 200 & 301 & 1757 \\ \hline
Oracle          & 0   & 2   & 1   & 0   & 1   & 0   & 0   & 1   & 335 \\ \hline
Adobe           & 7   & 17  & 11  & 8   & 47  & 44  & 42  & 51  & 637 \\ \hline
ICS-CERT        & 92  & 90  & 143 & 123 & 183 & 205 & 208 & 184 & 1073 \\ \hline
MITRE           & 196 & 153 & 154 & 129 & 353 & 184 & 216 & 218 & 1011 \\ \hline
IBM             & 56  & 219 & 234 & 48  & 53  & 132 & 105 & 103 & 840 \\ \hline
Qualcomm        & 19  & 34  & 378 & 3   & 49  & 66  & 62  & 129 & 829 \\ \hline
Samsung M.      & 97  & 155 & 452 & 132 & 102 & 259 & 281 & 320 & 872 \\ \hline
Dell            & 38  & 204 & 98  & 106 & 154 & 166 & 155 & 217 & 872 \\ \hline
GitLab          & 38  & 197 & 137 & 118 & 116 & 172 & 169 & 147 & 834 \\ \hline
Intel           & 1   & 271 & 52  & 210 & 167 & 94  & 69  & 96  & 540 \\ \hline
Red Hat         & 13  & 80  & 76  & 33  & 42  & 40  & 55  & 48  & 512 \\ \hline
Siemens         & 26  & 48  & 58  & 18  & 51  & 55  & 45  & 71  & 491 \\ \hline
Snyk            & 16  & 125 & 44  & 32  & 62  & 153 & 153 & 184 & 547 \\ \hline

\end{tabular}
\end{table}

\begin{figure*}
\centering
\includegraphics[width=\linewidth]{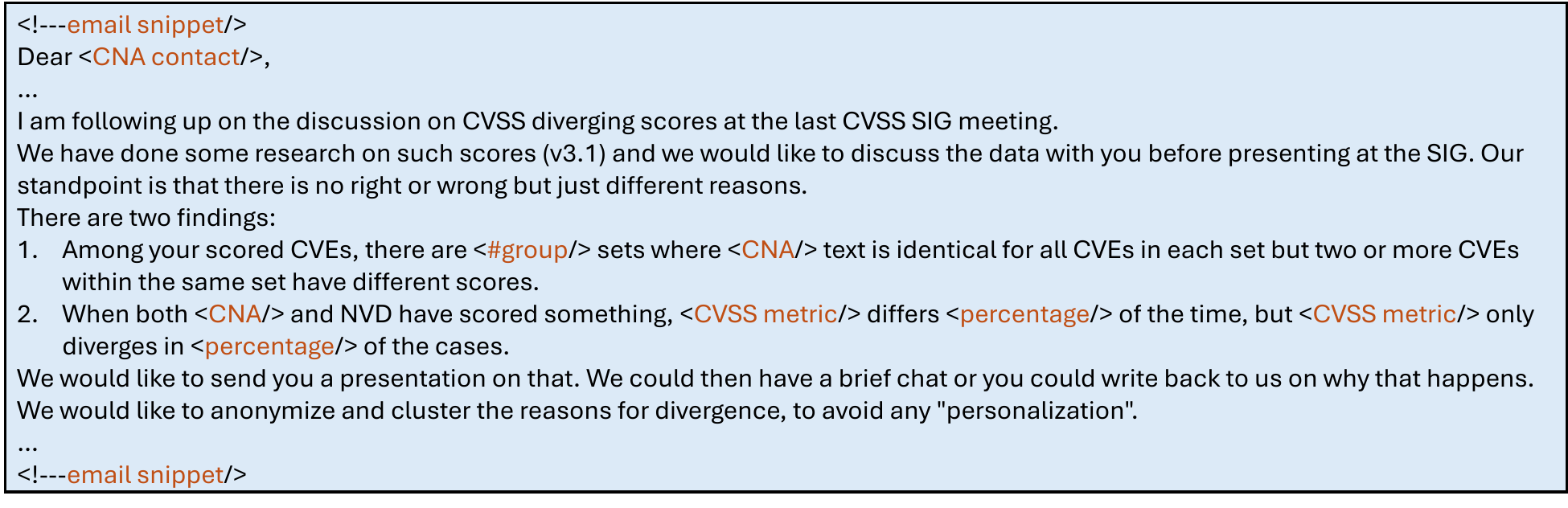}
\caption{First Email: Initial email inviting CNA representatives to review our preliminary findings}
\label{fig:first_email}
\end{figure*}

\begin{figure*}
\centering
\includegraphics[width=\linewidth]{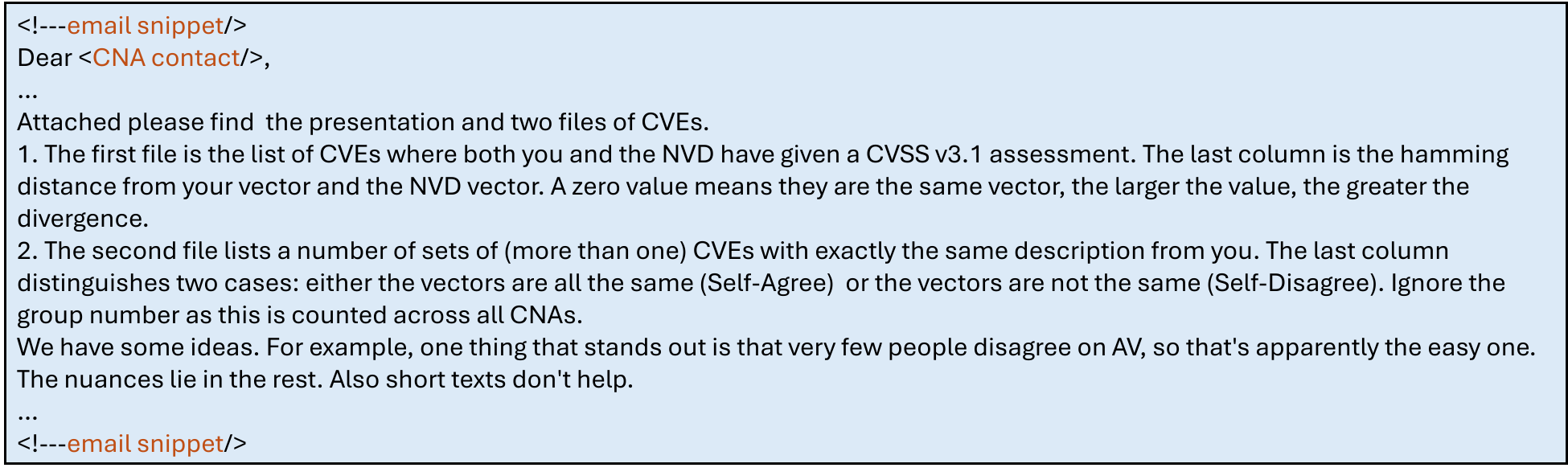}
\caption{Second Email: Follow-up email sharing the presentation slides and CNA-specific analysis for feedback and discussion}
\label{fig:second_email}
\end{figure*}

\clearpage

\section{Experiment} \label{sub:expAppendix}

\begin{figure*}[h]
\centering
\footnotesize
\includegraphics[width=\linewidth]{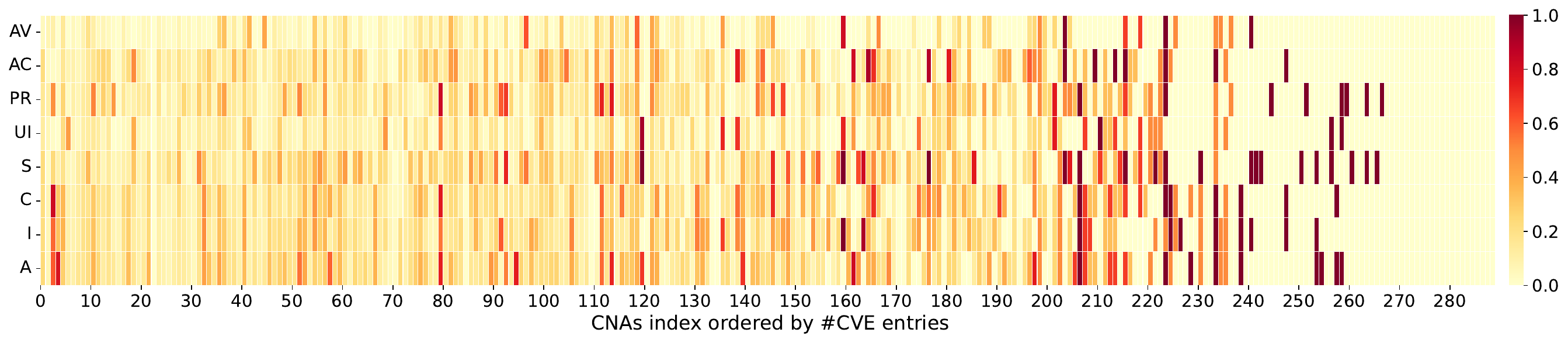}
\caption{Heatmap of all 266 public CNAs}
\label{fig:266CNAs_pmdc}
\end{figure*}

\begin{figure*}[h]
\centering
\footnotesize
\includegraphics[width=\linewidth]{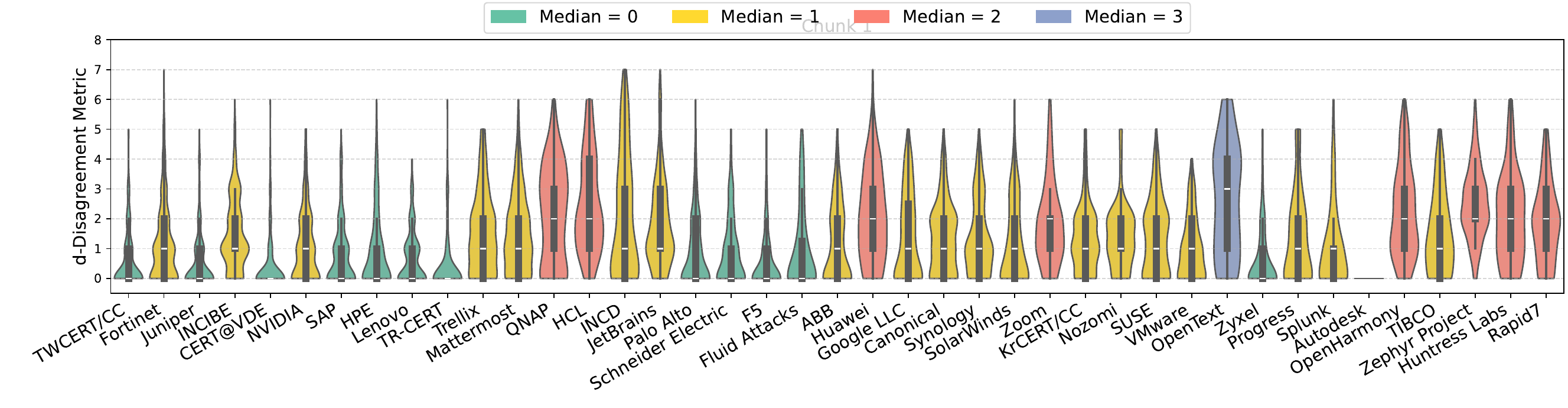}
\caption{Violin plot of the rest public CNAs}
\label{fig:next1CNAs_violin}
\end{figure*}

\begin{figure*}[h]
\centering
\footnotesize
\includegraphics[width=\linewidth]{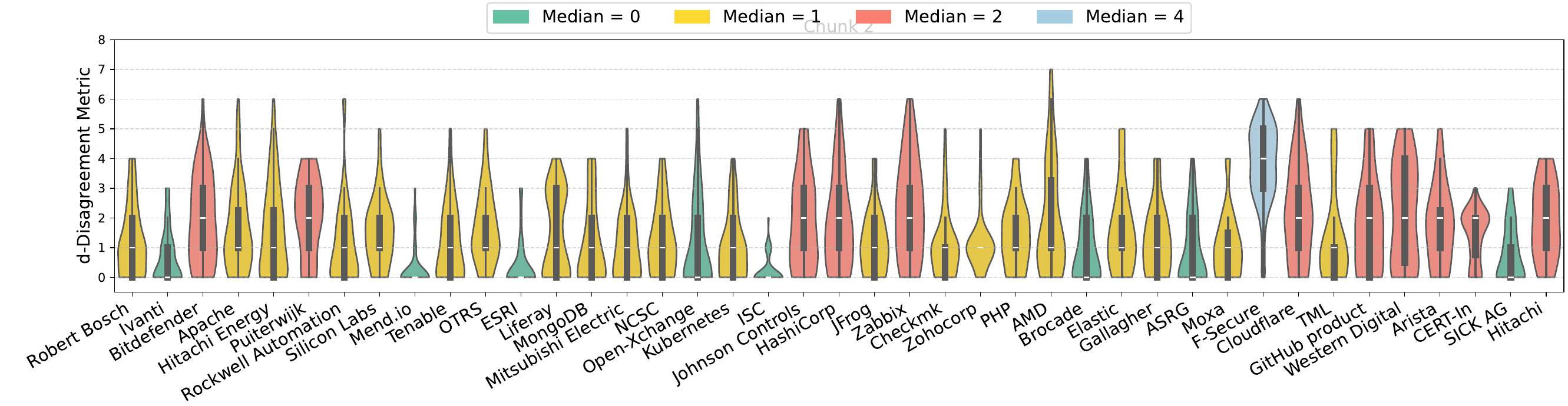}
\caption{Violin plot of the rest public CNAs}
\label{fig:next2_voilin}
\end{figure*}

\begin{figure*}[h]
\centering
\footnotesize
\includegraphics[width=\linewidth]{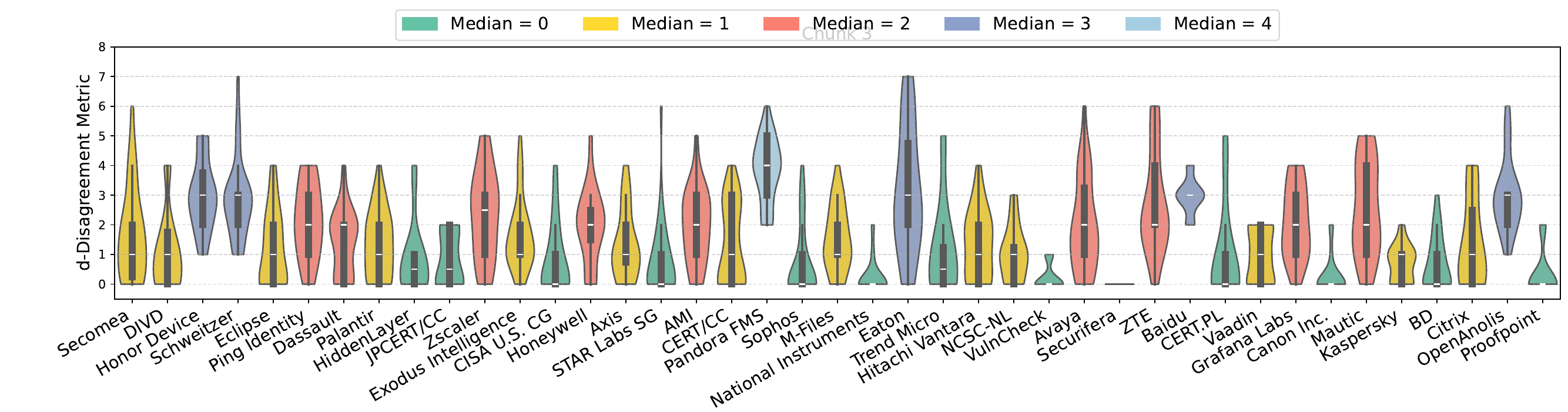}
\caption{Violin plot of the rest public CNAs}
\label{fig:next3_voilin}
\end{figure*}

\begin{figure*}[h]
\centering
\footnotesize
\includegraphics[width=\linewidth]{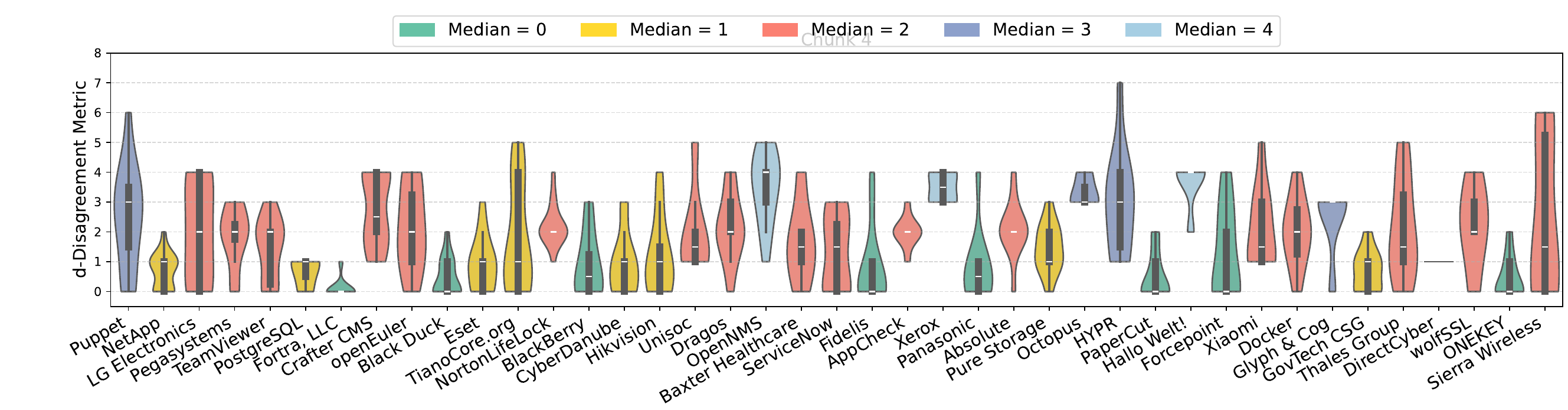}
\caption{Violin plot of the rest public CNAs}
\label{fig:next4_voilin}
\end{figure*}

\begin{figure*}[h]
\centering
\footnotesize
\includegraphics[width=\linewidth]{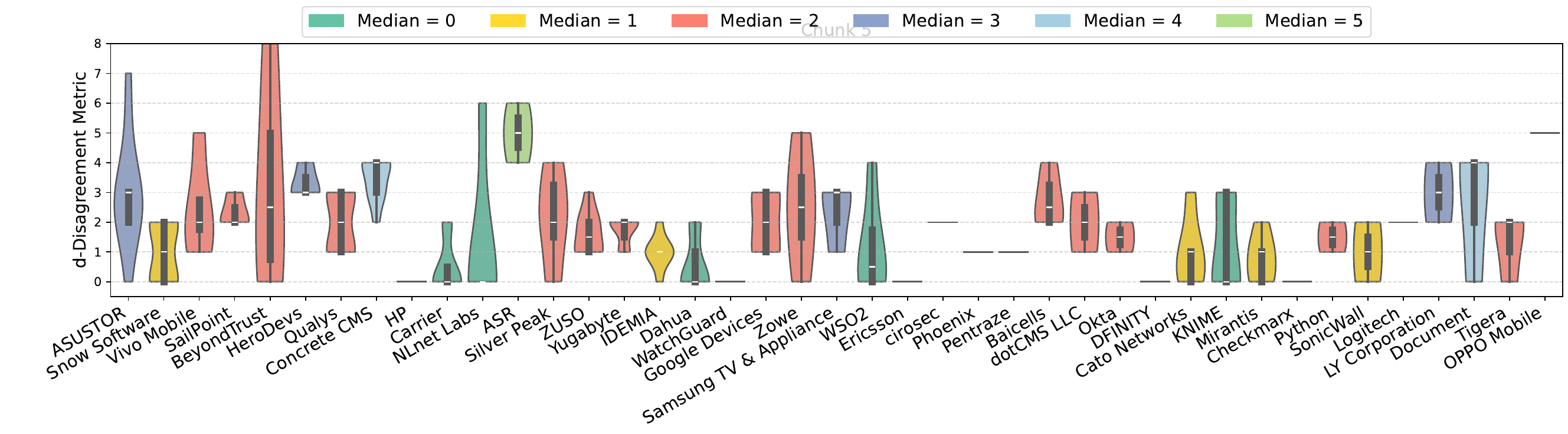}
\caption{Violin plot of the rest public CNAs}
\label{fig:next5_voilin}
\end{figure*}

\begin{figure*}[h]
\centering
\footnotesize
\includegraphics[width=\linewidth]{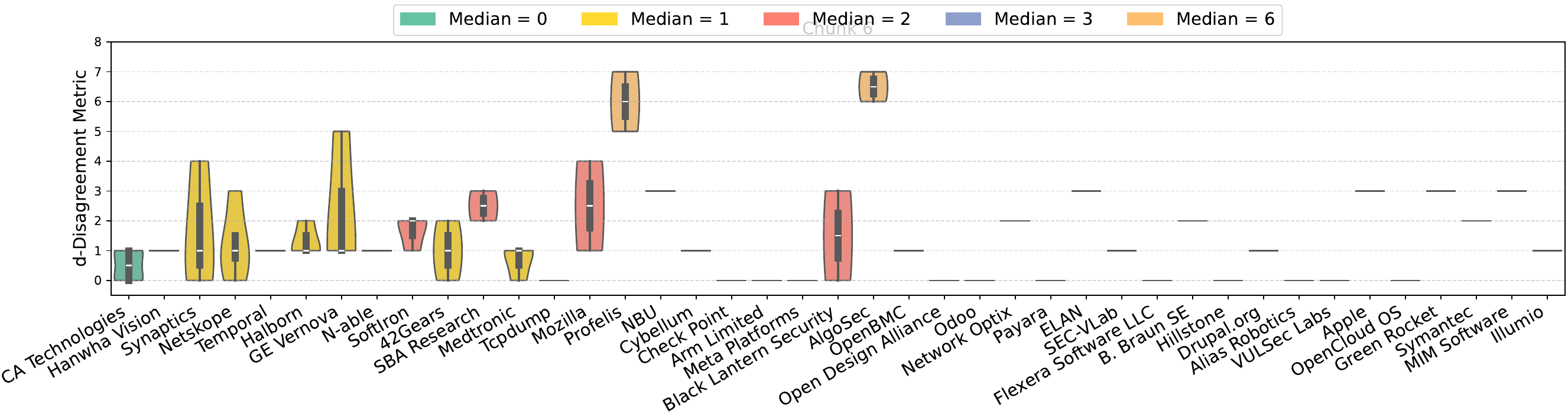}
\caption{Violin plot of the rest public CNAs}
\label{fig:next6_voilin}
\end{figure*}

\end{document}